# Tailoring Structure-borne Sound Through Bandgap Engineering in Phononic Crystals and Metamaterials: A Comprehensive Review

Mourad Oudich[1,2,*,†], Nikhil JRK Gerard[1,†], Yuanchen Deng[1], and Yun Jing[1,‡]

[1]Graduate Program in Acoustics, The Pennsylvania State University, University Park, Pennsylvania 16802, USA

[2]Université de Lorraine, CNRS, Institut Jean Lamour, F-54000 Nancy, France

* mxo5236@psu.edu

‡ yqj5201@psu.edu

† MO and N JRK G contributed equally to the present work.

**Abstract**

In solid state physics, a bandgap (BG) refers to a range of energies where no electronic states can exist. This concept was extended to classical waves, spawning the entire fields of photonic and phononic crystals where BGs are frequency (or wavelength) intervals where wave propagation is prohibited. For elastic waves, BGs were found in periodically alternating mechanical properties (i.e., stiffness and density). This gave birth to phononic crystals and later elastic metamaterials that have enabled unprecedented functionalities for a wide range of applications. Planar metamaterials were built for vibration shielding, while a myriad of works focused on integrating phononic crystals in micro-systems for filtering, waveguiding, and dynamical strain energy confinement in optomechanical systems. Furthermore, the past decade has witnessed the rise of topological insulators which lead to the creation of elastodynamic analogs of topological insulators for robust manipulation of mechanical waves. Meanwhile, additive manufacturing has enabled the realization of three-dimensional (3D) architected elastic metamaterials which extended their functionalities. This review aims to comprehensively delineate the rich physical background and the state-of-the art in elastic metamaterials and phononic crystals that possess engineered BGs for different functionalities and applications, and to provide a roadmap for future directions of these manmade materials.

# 1. Introduction

The concept of BGs was first introduced for classical waves in 1987 when Yablonovitch and John [1,2] demonstrated that a periodic arrangement of materials with contrasting optical indices could yield frequency intervals (BGs) that were forbidden to electromagnetic wave propagation. These BGs are analogous to those found in the energy band structure for crystalline materials. They can be characterized by evaluating the frequency of the propagating modes as a function of the wave vector magnitude in the different propagation directions along the Brillouin zone associated with the periodic configuration of the lattice. The optical structure that was employed for this realization was coined photonic crystal (PtC) and has ever since served as the cornerstone for modern research in wave physics and engineering optical devices [3,4]. Inspired by PtCs, phononic crystals (PnC) were subsequently introduced by Tamura, Hurley, and Wolfe [5] in 1988 for elastic waves in a one-dimensional (1D) periodic structure, which was followed by the pioneering studies of Sigalas and Economou [6,7] and Kushwaha $et\ al.$ [8,9] in early $90^{th}$ where the existence of phononic BGs was demonstrated for elastic waves. A PnC is made of elastic or fluid units (or scatterers) that are periodically distributed in a host medium (either elastic or fluid) that has contrasting mechanical properties (density and elastic modulus or compressibility). The two decades following the introduction of PnCs have witnessed countless studies that put forward and examined diverse PnC designs [8–30] alongside proposing modeling approaches specifically constructed to characterize their wave dispersion through numerical techniques like Plane Waves Expansion [8,9,11–13,18,19,21–23,26,27,29,30], Finite Difference Time Domain [16,20,24], Multiple Scattering Theory [14,17], and Finite Elements methods [25,26,28]. The primary objective of these early works was centered around the physics of mechanical wave dispersion in PnCs with the goal of unveiling the mechanisms behind the opening of the BGs. These works thus studied the influence of geometry, constituent materials, and periodicity on the BGs opening, their localization and width within the band structure. These methods thus not only facilitated the exploration of a large set of designs with structural complexity, but also helped demonstrate novel and remarkable mechanical wave phenomena such as slowing down the wave's group velocity [31,32], confining acoustic waves at structural defects [11,33,34], waveguiding [20,24,26,35–39], and even setting up preferential directional wave propagation [40]. Nowadays, phononic BG holds an important place in modern research on wave physics and engineering mechanical wave-based devices and constitutes a fundamental means for the realization of wave confinement and filtering that are essential for high device performance in the fields of optomechanics, energy harvesting, mechanical vibrations shielding, and robust topological wave transport.

In the early PnC studies, the opening of the BG was attributed to the Bragg scattering mechanism caused by the periodicity of the lattice. Tamura $et\ al.$ [5] were the first to explore the mechanism of acoustic Bragg reflection using a 1D PnC which displayed a barrier functionality for both longitudinal and transverse elastic waves at specific frequency intervals. Afterwards, numerous studies were conducted to evidence the Bragg BG in different phononic structures while exploring their dispersion. It was then well established that Bragg BGs strongly depend on the periodicity and the symmetry of the PnC, and that the operating wavelengths in the BG are on the order of magnitude of the periodicity. Meanwhile, in some other cases, BGs also may originate from another mechanism far different from Bragg scattering. Highlighted for the first time by Liu $et\ al.$ [41] in 2000, this mechanism is generally associated with strong localized elastic resonances within the PnC which was labeled as locally resonant PnC (LRPC). In the structure proposed by Liu $et\ al.$ [41], the units of periodicity are spherical resonators with low resonance frequencies and the BG is created from the coupling between the resonance of these spheres and the acoustic modes that propagates inside the PnC. Hence, the BG strongly depends on the mechanical properties of individual resonant units rather than the periodicity of the structure. Further, the resonators can be tailored to make the BG appear at wavelengths significantly larger than the periodicity of the LRPC which, hence, allows for the application of the homogenization theory to extract the dynamic effective properties of the structure. Under this homogenization assumption, it was revealed that the LRPC displays highly divergent negative effective mass density and/or elastic modulus at the BG frequency range [41–58]. These LRPCs are also known as metamaterials, and their highly unconventional effective properties have attracted great attention within the communities working on PnCs at that time, and led to the proliferation of subsequent works that explored the effective dynamic properties of LRPCs, thereby giving birth to the sub-fields of acoustic and elastic metamaterials.

The concept of a metamaterial was originally introduced by Veselago in 1968 for optical waves as the theory for a prospective functional materials that could exhibit negative refractive index [59]. But the absence of natural materials and means for the realization of a demonstrator with such abnormal property have hindered the interest into this field. It was not until 1996 when Pendry *et al*. [60,61] proposed a realistic structure comprised of metal wires and displayed a negative effective permittivity, and another structure made of two concentric split-ring-resonators with negative effective magnetic permeability [62]. Then, Smith *et al*. [63] first constructed an optical structure capable of manifesting simultaneous negative electrical permittivity and magnetic permeability, which led to the first experimental demonstration of negative optical refraction [64]. This realization had sparked immense interest towards this kind of artificial materials with negative optical properties, called electromagnetic metamaterials. The structure proposed by Smith *et al*. [63] was made of a periodic distribution of resonators well-tailored to endow the structure with negative effective permittivity and/or permeability at certain frequency ranges [60–65]. This led to the realization of optical devices with unconventional capabilities such as negative-index materials [66,67], optical super-lenses [65,68], cloaking [69,70], wavelength demultiplexing [71], and reversed Doppler effects [72,73]. These outstanding efforts inspired parallel works that went on to demonstrate analogous functionalities for mechanical waves. Acoustic metamaterials were then constructed as periodic structures with mechanical resonators well designed to manifest negative effective mechanical properties, i.e., density and compressibility in the case of fluid-borne sound or bulk modulus in the case of structure-borne sound. Intuitively, the negativity is associated with the material's abnormal response to dynamic solicitations as it tends to expand for a compressional pressure in the case of negative compressibility, while its acceleration becomes opposite to the applied force for the case of negative density. These special behaviors reshaped and propelled the field of mechanical waves over the past two decades and enabled exotic functionalities like sub-wavelength sound mitigation [74,75], negative refraction [76], ultrasound superfocusing [77], and acoustic cloaking [78].

PnCs can be constructed through multiple possibilities of periodic structuration with different geometrical shapes and materials for the constituent units. For elastic waves, the design of PnC strongly depends on the kind of wave that is of interest. Early studies in PnCs mainly focused on a two-dimensional (2D) periodic distribution of infinite cylindrical inclusions in a hosting medium using isotropic materials. This design configuration not only presents less complexity in terms of the couplings between elastic modes, but also offers simplicity in solving the governing equations accommodating the limited computational capabilities at that time [7–9,11,20,29,36,37,79]. Three-dimensional (3D) periodic structures were also considered by adopting atomic-like arrangement (simple cubic, body-centered cubic, and face-centered cubic) of spherical inclusions in a host medium [6,10,16,17,41,80–87], and efficient computational methods were introduced to precisely characterize the wave dispersion through the calculation of the band structure. Some studies also focused on using PnCs for the manipulation of surface acoustic waves (SAW) [12,13,21,22,24,38,88–106] either in the microscale for SAW filtering [89,91–93,95,96,98,102–104] or for seismic waves shielding [97,99–101,105,106], while other works drove their interest towards the manipulation of Lamb waves in plates whose wavelengths are in the order of the plate's thickness [18,23,25–28,30,35,38,107–116]. Meanwhile, the advances in additive manufacturing techniques have expanded the geometrical and material design space of elastic metamaterials and have facilitated the fabrication of lightweight and mechanically strong structures with great potential for wave propagation applications. In fact, architected materials with precise structural building based on connected rods and masses have enabled multiple 3D structures manifesting ultra-wide BG for mechanical waves [117–129]. These efforts paved the way to a new generation of architected elastic metamaterials for omnidirectional wide BG while being lightweight with high porosity.

In the last decade, elastic BG has become a fundamental knowledge for both understanding wave physics and creating high performance devices for mechanical wave control. In optomechanics, phoxonic crystals (PxC) which are dual PnCs and PtCs, have become a vital platform for efficient control of both electromagnetic waves and elastic waves through engineering simultaneous optical and elastic BGs [130–151]. BG engineering has also been utilized to harvest mechanical energy carried by sound/vibration using a planar metamaterial endowed with subwavelength BG, and where a structural defect is introduced to allow for the existence of mechanical cavity modes at frequencies inside the BG [152–160]. For example, when an impinging sound wave reaches the metamaterial panel, it excites the phononic cavity modes with highly confined strain energy density, its energy can then be harvested using a piezoelectric material inside the defect. Moreover, PnCs have also served as an exotic platform for mimicking quantum phenomena in condensed matter physics, especially since the discovery of topological insulators. For elastic waves, these analogs includes Landau Level [161], quantum Hall based structures with robust chiral edge states [162], quantum spin Hall effect [163], elastic Valley-Hall edge states [164,165], and elastic higher-order topological insulator with topologically protected corner states [166]. These discoveries have enabled the creation of elastodynamic platforms where wave manipulation can be performed in a highly unusual but robust manner, with backscattering-free waveguiding that is immune to structural defects and fabrication imperfections. **Figure 1** presents a timeline for phononic and metamaterial platforms hosting elastic BGs that were engineered for exotic wave functionalities.

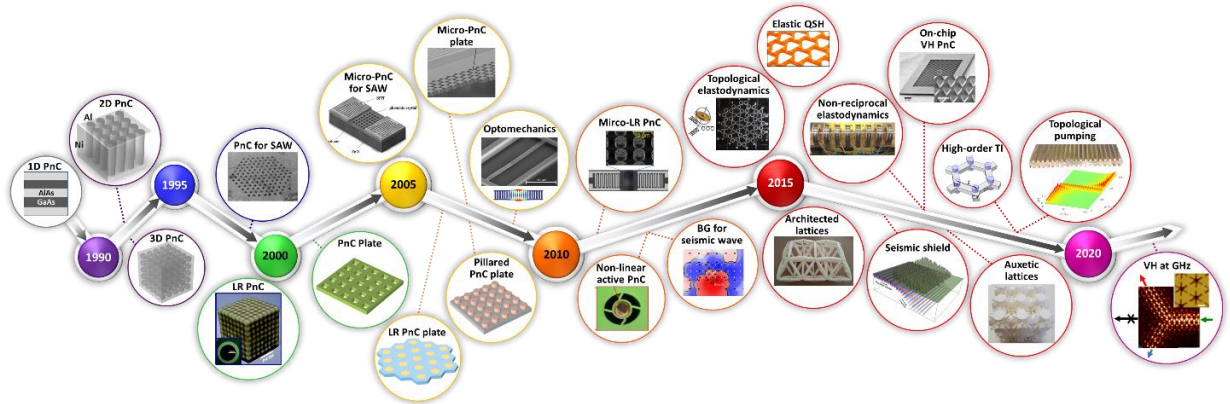

**Fig.1. Timeline for some milestone elastodynamic platforms endowed with elastic BG**. These works include the emergence of phononic crystals and elastic metamaterials, optomechanical crystals, nonlinear and active metamaterial, topological elastic lattices, 3D architected lightweight metamaterials, non-reciprocal elastic metamaterial, and topological pumping. PnC: Phononic Crystal; LR: Local resonance; BG: bandgap; QSH: Quantum Spin Hall; TI: topological insulator; VH: Valley Hall.

In this review, we browse through the establishment of the elastic BG in PnCs and elastic metamaterials and reveal its importance in wave physics and device engineering for wave manipulation applications spanning Hz to GHz frequency regimes. Here, we mainly focus on structure-borne sound (i.e., elastic waves) and specifically the physics of elastodynamic bandgaps and their applications. The term "structure-borne sound" is used to indicate sound that propagates in solid elastic materials, namely elastic waves. Compared to acoustic functional materials for controlling air-borne sound, our review paper is motivated by the fact that elastic wave functional materials cater to an entirely different set of applications such as SAW-devices, optomechanics, plate-type devices, seismic shielding, etc. For recent review papers on acoustic functional materials for controlling air-borne sound, the reader may refer to [167–169]. The second section of this review paper describes the main historical works on PnCs and metamaterials with focus on the physics of BG opening. Then, section 3 discusses the important achievements on the structural design of PnCs and elastic metamaterials while summarizing different routes for enlarging the elastic BG for different application purposes. We particularly deal with the proposed 3D designs in literature and their architectural building enabled by advanced additive manufacturing techniques. In section 4, we focus on the importance of BG for specific applications that encompass optomechanics, topological elastodynamics, energy harvesting, sensing, and active metamaterials for either frequency tuning of the wave filtering capability or to realize non-reciprocal wave propagation. Further, we reveal the numerical inverse design methods and optimization approaches for BG engineering, with focus on topology optimization and machine learning algorithms. We finalize our review on presenting an outlook on future routes for elastic BG applications. **Figure 2** gives the presented sections in this manuscript.

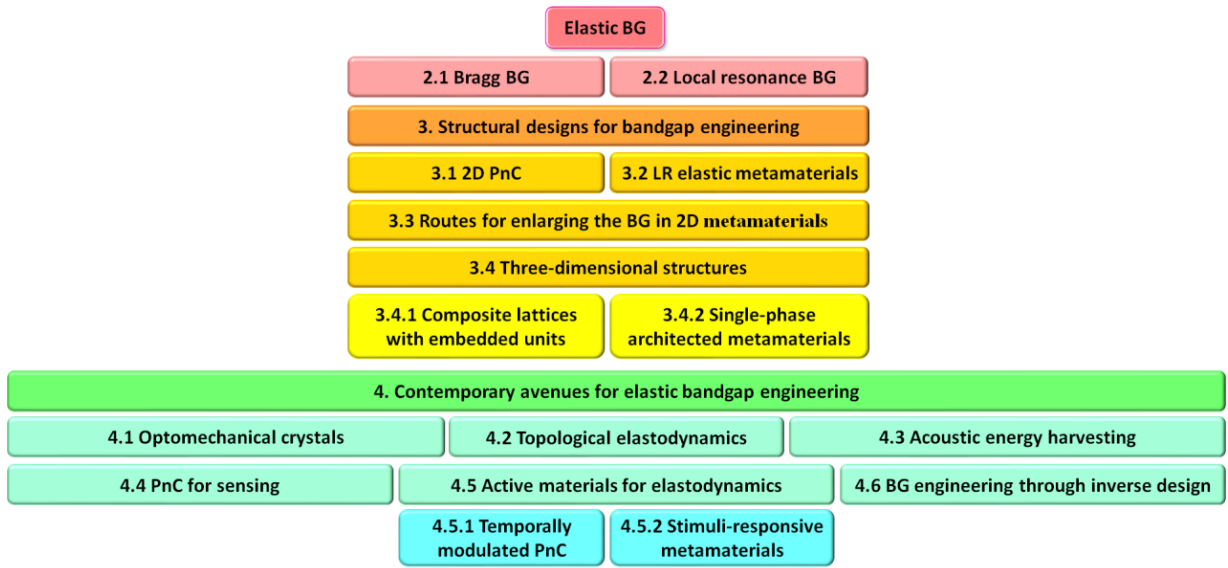

**Fig.2**. **Flow chart presenting the sections of this paper.** The main body of the paper is divided into three main sections: Section 2 concerns the mechanism of BG opening; section 3 presents the structural designs for BG engineering; section 4 gives an overview on application avenues for elastic BG.

## 2. Fundamentals of elastodynamic bandgaps

The now long-standing interest in PnCs mainly emanates from their capability of producing BGs for acoustic and elastic waves. The interaction of the wave with the periodic structuration of the crystal causes internal wave reflections and interferences or resonances that lead to evanescent waves with spatial exponential decays as the wave propagates through the PnC. Since the birth of PnCs, extensive works have been devoted to understanding the underlaying physics behind the opening of BGs that later led to the building of numerous phononic structures and multiple modeling approaches for their designs and characterization. Based on the underlying mechanisms, BGs can be broadly classified as those that occur as a result of Bragg scattering or/and local resonances. The origin and features of these BGs are introduced in the following section.

## 2.1 Bragg bandgap

Historically, early works on PnCs focused on BGs that are created from the scattering of elastic wave by the periodic structuration of the PnC [5–9]. PnCs were constructed by considering periodic layers in 1D propagation [5] or solid inclusions in a hosting medium in 2D [6–9] with highly contrasting elastic properties between the inclusions and the medium, which is key for the creation of the BG. The physical mechanism of BG opening was attributed to wave interference known as Bragg scattering that arises from the wave interaction with successive periodic rows of inclusions inside the crystal. Consequently, the Bragg BG occurs at frequencies where the wavelength is on the order of magnitude of the periodicity, specifically when half of the wavelength closely matches the periodicity (bottom right panel in **Fig. 3(a)**). Examples of applications are discussed in section 4. The existence of phononic BG was initially evidenced theoretically for elastic waves [5–8]. The wave dispersion was characterized by calculating the band structure analog to the electronic band structure where instead of energy, the frequencies of the propagating modes are evaluated as function of the wave vector amplitude in the Brillouin zone associated with the lattice. An omnidirectional phononic BG is identified as the frequency region where no eigenmode can exist in all directions of the irreducible Brillouin zone. During the 90s and beginning of the 21$^{st}$ century, acoustic and elastic band structures were investigated to understand the physical mechanism of Bragg BG in multiple PnC designs for different kinds of waves: bulk waves with 2D [6–9,15,170–173] or 3D periodicity [6,10,16,17,80,81,83,85,87,174–178], guided plate waves (Lamb waves) [18,23,25–28,30,35,38,107–116] (**Fig. 3(a)** upper middle panel), and SAW [12,13,21,22,24,88–91,94,96] (**Fig. 3(a)** upper right panel). Since these Bragg BGs are created due to the destructive interference caused by the periodicity inside the PnCs, the frequency of the BG is proportional to the size of the PnC. Most of these works characterized the wave dispersion by displaying the band structure with the normalized frequency $\omega a/v$ where $\omega$ is the angular frequency, $a$ is the lattice constant of the PnC, and $v$ is the smallest shear wave velocity of the bulk medium. Large structures can be designed for low frequency BG while building small size PnC (at the micro and nano scales) lead to BG operating at very high frequency range (hundreds of MHz to few GHz) as it will be detailed in **section 4** through some examples.

Meanwhile, if one creates a structural defect inside the PnC by changing either the geometry or the material composition of one scatterer or multiple scatterers along a line, elastic waves can be trapped or guided inside the defect where the wave energy becomes confined. These trapped or guided modes appear in the band structure as additional modes in the frequency range of the BG. The latter then became a fundamental physical tool for engineering phononic platforms for wave control for a wide variety of applications including filtering, waveguiding [20,24,26,35–37], demultiplexing [179], optomechanics [130–151], and acoustic energy harvesting [152–160]. The physics of Bragg BG opening was also investigated by calculating the complex band structure for evanescent waves which put forth the wave dispersion complexity inside the PnC, especially in the BG frequency range [29,30]. Interestingly, the last decade has seen the revival of interests for Bragg BGs after the discovery of acoustic analogs of topological insulators. The topological features of the bands with regards to the PnC crystal symmetry was investigated to create trivial and non-trivial BGs that could be used to create robust guided interface modes between two PnC with different topologies. Furthermore, and recently, the Bragg BGs were exploited to realize non-reciprocal acoustic and elastic wave propagation. The properties of the PnC were modulated both in space and time to break the time-reversal symmetry and realize unidirectional BGs. These topics will be discussed in further detail in **section 4.2**.

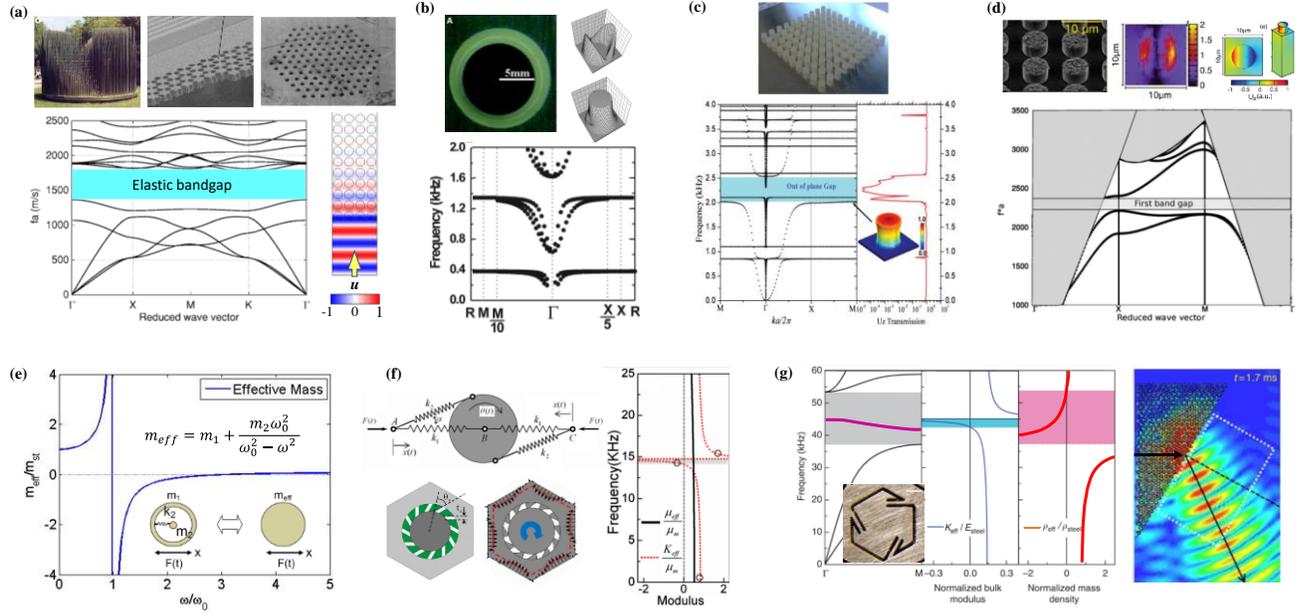

**Fig.3. Phononic crystals and locally resonant metamaterials.** (a) Examples of PnCs. (Top from left to right) 2D lattice of rods in air [170], micro holes in a plate [114], and holes in a mable quarry [88]. (bottom panel) band structure of a PnC plate with a BG (highlighted in cyan) [25] and wave propagation at a frequency inside the BG. (b)-(d) Locally resonant PnC : (b) unit cell of a 3D lattice [41], (c) pillar resonators on a plate [180], ans (d) micro-pillared surface [95]. (e)-(g) Negative dynamic properties of elastic metamaterials: (e) negative mass [50], (f) negative bulk modulus [51], and (g) a double negative elastic metamaterial leading to negative refraction [55].

### 2.2 Local resonance bandgap

BGs can also be created using strong mechanical resonances as first demonstrated by Liu *et al.* [41] in 2000. Their proposed phononic structure is made of a 3D lattice of spherical resonant units distributed periodically in a simple cubic configuration embedded in a host matrix of epoxy. Each resonator is made of a heavy core of lead coated with a very soft material (silicone rubber) (**Fig. 3(b)**). The pair of lead core and silicone rubber coating behaves like a spring-mass resonator with low stiffness and heavy mass leading to resonance modes located at very low frequencies. The physical mechanism of the BG is based on the coupling between the resonance modes of the spherical units, which are localized, and the bulk modes inside the crystal. In the band structure, resonances appear as flat bands away from the $\Gamma$ point of the Brillouin zone (**Fig. 3(b)**) indicating zero group velocity, and delimiting the lower edge of the BG. Further, the local resonance (LR) based BG is governed by the geometry and mechanical properties of the resonator, and does not depend on the periodicity nor the symmetry of the crystal [181]. The choice of materials is key in this case as low stiffness materials included in the resonators are used to enable the LR in a stiffer matrix (epoxy). Consequently, the LR mechanism can be tailored to occur at low frequencies where the wavelengths in the hosting media can be several orders of magnitudes larger than the periodicity of the crystal (deeply subwavelength). In the structure designed by Liu *et al*. [41], the BG was found at 380 Hz (**Fig. 3(b)**) where the longitudinal wave wavelength in the epoxy matrix is around 300 times the periodicity. The deeply subwavelength BGs enabled by LRPC opened new routes for exploring the physics of wave dispersion at the microscopic scale towards the design of reduced size structures for wave shielding at the low audible frequency range. This also sparked the exploration of other 3D LRPC designs [43,84,182] alongside with 2D structures [183–187] based on soft inclusions in a stiff matrix. Other studies have proposed multiples elastic systems with LR BG for Lamb waves using plates with either soft inclusions [188,189] or resonant pillars [28,110,111] (**Fig. 3(c)**), and for SAW [93–95,190] (**Fig. 3(d)**).

Considering the deeply subwavelength functionality of the LRPC, homogenization methods were used to study their effective dynamic behavior which led to the demonstration of anomalous dynamic properties that have attracted significant attention. In fact, at the resonance frequencies where the wavelengths in the hosting media are larger than the periodicity of the structure, Liu *et al.* [41] showed that their structure manifests a divergent and negative effective mass density at the regions of the LR BG. This groundbreaking discovery has served as the cornerstone for acoustic and elastic metamaterials that have reshaped the paradigm of mechanical wave propagation. Multiple theoretical studies were conducted to investigate the negativity of the effective density in elastic metamaterials [43,46,48–50,53,54]. Furthermore, Li and Chan [42] demonstrated that acoustic metamaterials can be designed to display both negative effective mass density and bulk modulus using rubber spheres in water. Then, for elastodynamics, Milton and Willis [47] presented a generalized Newton second law to better describe the dynamics of LRPC especially in the extreme case of divergent effective mass density at the resonance. Afterwards, several studies demonstrated double negativity upon the effective mass density and the effective bulk and/or shear modulus in elastic metamaterials [44,45,51,52,55–58]. Analytical formulations were developed based on the spring-mass model [50,51] to understand the concepts of negative effective mass and bulk modulus and their relationship to the opening of the BG (**Fig.3 (e)-(g)**). Recent works conducted by Dong *et al*. [56–58] used topology optimization to design double negative elastic metamaterials. Single negative density or bulk modulus leads to LR BG opening that causes total wave reflection at the resonance with values of the effective parameter tending to be infinite in a lossless system. However, simultaneous negative density and elastic modulus at the same frequency range allows for the appearance of a propagating band with the particularity of endowing the structure with negative effective index, leading to negative refraction of the wave (**Fig.3 (g)**) [55,58]. The extensive works on these negative metamaterial behaviors [41–58] led to unprecedented wave functionalities such as sub-wavelength sound mitigation [74,75], super-focusing [77], and acoustic and elastic cloaking [56,78,191].

## 3. Structural designs for bandgap engineering

The opening of an elastic BG and the control of its width and localization within the band structure has long been a subject of concern for researchers working on the topics of PnCs and elastic metamaterials. BG engineering has thus surfaced due to its usefulness for applications like vibration mitigation, seismic shielding, multidirectional wave cancellation, SAW filtering, long-life time phonon confinement for optomechanics, and shaping the topology of the bands to harbor topologically robust waveguiding and confinement that are immune to defects. In this section, we present examples of PnCs and elastic metamaterials engineered for the purpose of BG manifestation while focusing on wave manipulations in 2D and 3D space for different kinds of guided waves associated with specific applications. The road of studying elastic BG and their utilization has led to a number of PnC and metamaterials designs, especially for the case of 2D lattices. Particularly, the control of guided waves such as Lamb waves and SAW, has led to considering structures with planar periodicity.

### 3.1 Two dimensional phononic crystals

Early studies on elastic BG were conducted on 2D periodic infinite inclusions, mostly of cylindrical shape, in a hosting medium for bulk waves dispersion [7–9,15,170,172,173]. The objective of these works was not only to demonstrate theoretically and experimentally that a periodic mismatch in the mechanical properties of the material creates BGs, but also to investigate the Bragg scattering mechanism that is behind their origin. The aforementioned studies on 2D PnCs focused on elastic waves using solid inclusions in a solid matrix. Kushwaha *et al.* [8] adopted both Al inclusions in Ni and Ni inclusions in Al, but only considered the dispersion of pure shear waves with out-of-plane polarization (displacement along the axis of the cylinders). Shortly after, a first complete BG was evidenced by Sigalas and Economou using Au inclusions in Be [7]. Although the tools for experimental demonstration were easily accessible in acoustics using solid rods in air [170,172], BGs for elastic waves remained purely theoretical for quite a while since fabricating a prototype comprising solid inclusions in a solid matrix was challenging. It was not until 1998 when the first experimental realization of an elastic PnC was presented by Montero de Espinosa *et al.* [171]. They revealed the existence of a BG for longitudinal waves in PnC made of a thick aluminum plate with holes filled with mercury. The mercury was chosen to create high stiffness and density mismatches with aluminum while the values of the acoustic impedances are close between the two materials in order to reduce mode conversions from the wave scattering inside the structure. Shortly after, García-Pablos *et al.* [15] built a PnC made of an aluminum slab with periodic cylindrical holes filled with either air, mercury or oil. In 2001, Vasseur *et al.* [173] fabricated a composite PnC made of a triangular array of steel cylinders in epoxy and showed for the first time the existence of an absolute BG for bulk waves, independently from the direction of propagation. To gain more insight on the Bragg BG opening, Laude *et al.* [29] and Moiseyenko and Laude [192] developed a plane waves expansion method to calculate the complex band structure of 2D PnC made of infinite cylindrical inclusions in hosting material. They were able to precisely characterize the dispersion of evanescent waves inside the PnC by evaluation the complex value of the wave number in the frequency regions of the BGs for each elastic polarization. Veres *et al.* [193] also investigated the evanescent waves dispersion in 2D PnC made of square lattice of holes with different shapes.

Regarding SAW, Meseguer *et al.* [88] experimentally demonstrated for the first time the existence of BG for Rayleigh waves. They considered hexagonal and honeycomb lattices of drilled cylindrical holes on the surface of marble quarry (**Fig. 3(a)**, upper right panel). In the same year, Tanaka and Tamura [13] investigated theoretically the BG opening for Rayleigh waves using a triangular lattices of cylinders embedded in a background media. These works sparked growing interest on integrating PnCs into microelectromechanical systems and SAW devices. Wu *et al.* [89] initiated this micro-integration by fabricating a PnC at the microscale made of a square lattice of holes in a silicon substrate. They used slanted interdigital transducers on top of ZnO layers to generate and detect the SAW via the inverse and direct piezoelectric effects and demonstrated the existence of BG at 200 MHz. In another study, Benchabane *et al.* [91] fabricated a piezoelectric PnC made of air holes etched in lithium niobate substrate and measured the Rayleigh waves attenuation associated with the BG extending from 203 to 226 MHz. Later on, an ultrafast optical technique was developed [194–196] to track in real time the SAW interaction with a PnC in a silicon substrate, allowing a first experimental extraction of the band structure. In another study, Liu *et al.* [96] designed and fabricated a PnC with BG for Love waves. Their structure was made of a square lattice of holes etched on a silica guiding layer deposited on the surface of quartz substrate. Meanwhile, PnCs crystals were also considered at larger scale for seismic waves shielding. Brûlé *et al.* [97] conducted a first large scale experiment on Rayleigh waves scattering caused by holes dug in soil and demonstrated the existence of partial BGs for seismic waves.

Phononic Bragg BGs were also demonstrated for Lamb waves using engineered PnC plates. Wilm *et al.* [18] proposed PnC plates with 1D and 2D periodicities while considering the piezoelectric effect using quartz inclusions in the composition of the PnC. Later, Hsu and Wu [23] and Khelif *et al.* [25] have theoretically investigated the dispersion of Lamb waves in plates with periodic cylindrical inclusions to show that a full BG can be created in the whole Brillouin zone. Afterwards, Hsiao *et al.* [107] experimentally exposed the BG in a PnC slab made of spherical beads in a solid epoxy matrix. Bonello *et al.* [108] designed and fabricated a PnC plate at the microscale deposited at the surface of a silicon plate. It was made of cylindrical iron inclusions embedded into a copper background, where the authors experimentally characterized the band structure for the first time and showed that depending on the filling fraction, frequency gaps appear for the lowest-order symmetric and antisymmetric modes. At the same time, Morvan *et al.* [109] studied Lamb waves interaction with a phononic micro-grating and experimentally evidenced the BG at the ultrasonic regime. Shortly after, Djafari-Rouhani *et al.* [38] and Vasseur *et al.* [26] optimized the BGs in PnC plates and leveraged them to achieve waveguiding by creating a defect line inside the phononic slab. This line defect was introduced by creating a wide geometrical spacing between the cylindrical units along specific directions of the periodicity. In another study, Sun and Wu [113] used the FDTD method to study wave propagation in a 2D phononic plate consisting of cylindrical steel inclusions. With the existence of the BG, they achieved waveguiding by reducing the size of the steel inclusions along a line. Besides, it became obvious to used Lamb waves in microsystems for filtering since they have the advantage of being well confined within the plate, contrary to Rayleigh waves. In fact, the scattering of Rayleigh waves by PnC generates bulk waves that propagates into the substrate, which significantly reduces the amplitude of the transmitted surface wave through the PnC outside the BG regions. Mohammadi *et al.* [114] designed a PnC microstructure made of hexagonal lattice of holes in a silicon plate using micro-fabrication process based on semi-conductor technology (**Fig. 3(a)** upper-central panel). The transmission measurements through the micro-PnC displayed high wave attenuation in the region of the theoretically predicted BG at 134 MHz with a bandwidth of 23%. Evanescent Lamb waves were also characterized by Oudich and Assouar [30] in phononic plates with inclusions made of different sets of materials. Other designs considered cross-like holes in a plate and showed that they can give rise to multiple complete BGs [115,197]. Miniaci *et al.* [116] proposed a phononic plate made of unit cells with cross holes at different hierarchical scales and evidenced the existence of BGs at multiple frequency ranges associated with the size of the constituent unit cells of each hierarchical order.

The interest in Bragg BG engineering has grown fast in the last decade thanks to its advantages for high performance optomechanical devices, alongside the realization of non-reciprocal wave propagation, and the creation acoustic and elastic states that are only governed by the topology of the bands. These subjects will be addressed in **section 4.**

### 3.2 Locally resonant elastic metamaterials

After the discovery and introduction of LRPC by Liu *et al.* [41], multiple studies focused on 2D ternary designs to reproduce the LR based BG at the subwavelength scale. Goffaux *et al.* [183,184] used the FDTD method to analyze the transmission spectrum of elastic waves through a lattice of cylindrical cores coated with a soft material, and investigated the effect of different materials of the core on the resonance. The LR BG was depicted in the transmission spectrum in the form of sharp asymmetric Fano like resonance. Similar ternary 2D systems were investigated using analytical and numerical methods [43,181,186,198]. Furthermore, Sheng *et al.* [181] considered non-periodic arrangement of inclusions to highlight the non-dependence of the LR BG on the periodicity of the structure.

Meanwhile, it is not necessary to consider ternary materials in the phononic structure to access LR BG. Wang *et al.* [185] introduced for the first time a binary LR PC made of cylindrical inclusions of soft rubber in an epoxy host. The key for the creation of LR BG is the high contrast in the mass density and elastic moduli between the soft material of the resonant units and the stiff material of the hosting medium. However, as the resonance phenomena is generally narrow band, it was challenging to produce the LR BG with a relatively wide frequency range which is of great interest for sound and vibration shielding applications. An attempt for the enlargement of LR BG was conducted by Larabi *et al.* [187] who proposed a structural design made of coaxial multilayered cylindrical inclusions of alternating soft (rubber) and stiff materials (steel). They observed several resonance peaks in the transmission spectrum that were leveraged to further enlarge the LR BG.

Meanwhile, LR BGs were also evidenced for Lamb waves by Hsu and Wu [188] who used rubber cylindrical inclusions of square and hexagonal lattices in an epoxy plate. However, an experimental realization of a prototype made of embedded units in plates was challenging at that time. A more feasible design was proposed by Wu *et al.* [28] and Pennec *et al.* [110] who introduced for the first time a structure made of a plate decorated with square lattice of pillars to create LR BG from the coupling between the Lamb waves and individual resonances of the pillars. Then, Oudich *et al.* [111,180,199] adopted the same design to theoretically and experimentally demonstrate ultra-low frequency LR BG using homogeneous and composite cylindrical pillars made of soft rubber with lead cape, distributed on an aluminum plate (**Fig. 3(c)**). The design of plates decorated with pillars has extended the geometrical parameter space of the resonators for tailoring the LR phenomena which offers better control over the BG width and frequency localization. It was found that the position of the BG is significantly affected by the height of the pillars and the mass of the heavy cape, while its width mainly depends on the pillar's radius [111]. Furthermore, in addition to the employed materials, the shape of the pillars and their contact with the plate is of great importance for the LR-wave coupling which hence directly affects the position and the width of the BG. This was demonstrated using pillars with neck [200,201], or having conical [202,203] or hollow shapes [204], or multi-concentric building design [205]. Consequently, metamaterial plates decorated with pillars are of great potential for real word application for both sound and vibration mitigation [74,75,189,206–208].

Besides, LR BG were also engineered for SAW both for microsystems integration, and for large scale application to shield seismic waves. Khelif *et al.* [93] was the first to theoretically investigate Rayleigh waves dispersion by a square lattice of pillars decorating the surface of a silicon substrate. They demonstrated the existence of LR BG for Rayleigh waves, created from the resonance of the pillars under the sound line. In general, Rayleigh waves interaction with periodic structures often results in mode conversion to the bulk substrate. In the case of periodic holes with sufficient depth, the scattering of Rayleigh wave with the holes causes the leaking of the strain energy carried by the surface wave into the bulk causing a quick decay of the transmitted surface wave amplitude. The phenomena can be depicted in the linear dispersion curves where the SAW-BAW couplings occur at frequencies above the sound line. Conversely, in the case of BG created by the resonance of the pillars, the couplings between the resonance modes and SAW occur below the sound line, which therefore causes less energy leakage into the bulk [94]. At the microscale, Robillard *et al.* [209] performed a first experimental observation of individual resonances of cubic metallic pillars decorating the surface of a substrate, by employing an ultrafast laser pump and a probing technique through the photoelastic mechanism. Afterwards, Achaoui *et al.* [95] used an optical measurement technique to characterize the vibration of individual pillars interacting with Rayleigh waves, and observed the generated LR BG by a square lattice of cylindrical nickel pillars deposited on lithium niobate substrate (**Fig. 3(d)**). Using the same design, Yudistira *et al.* [98] showed both theoretically and experimentally the simultaneous existence of Bragg and LR BGs. Furthermore, LR BG were also exposed experimentally using a lattice of metallic nano-discs [92] and pillars in holes [102] for Rayleigh waves. Oudich *et al.* [103,104] designed phononic pillars with BG that were designed to host cavity modes through structural defects, and used these pillars to realize wave confinement, Fano resonances, and elastic analog of electromagnetically induced transparency for Rayleigh waves.

For large scale SAW, LRPC were designed to provide potential solutions for shielding seismic waves. Colombi et al. [99,100] introduced the idea of using a forest as a lattice where each tree behaves as a resonator with flexural vibration to strongly attenuate seismic waves [105]. Other designs of metamaterial made of either 2D lattice of resonators [101] or circular inclusions [106] in soil were proposed for seismic waves mitigation.

### 3.3 Routes for enlarging the bandgap in 2D metamaterials

The design of periodic structures for elastic BG engineering was stimulated by the growing interest for real world applications centered around shielding and filtering undesirable vibration, either in the ultrasonic or audible frequency ranges. Although the main advantage of LR-based BG is the access to small size structures operating at the subwavelength scale, the challenge of widening the BG was to be faced. This engaged a race towards engineering metamaterials that could shield the wave at the largest frequency range possible. The evident route for BG enlargement is acting upon the geometrical layout of the constituent materials and the contrast of their mechanical properties that strongly affects the wave dispersion. The first study dealing with BG engineering for the purpose of frequency width maximization was initiated by Yilmaz *et al.* [210,211]. They considered a design of connected masses in a particular configuration that is engineered to promotes the mechanism of inertia amplification. This mechanism allowed them to achieve extremely wide BG reaching a relative width (the frequency width of the BG divided by its central frequency) of 109%. Their theoretical studies were shortly followed by an experimental realization of an architected metamaterial made of 2D lattice of beams well-constructed and connected for the inertia amplification mechanism to operate [212].

Besides, multiple studies have delt with BG enlargement via design engineering by constructing numerical and optimization methods applied to 1D and 2D PnC, and the first works focused on lattices of unit voids in a solid material. Hussein *et al.* [213] built a multi-objective genetic algorithm to optimize the design of a 1D binary PnC in order to achieve maximum wave attenuation via broadband BG. For 2D phononic designs, Diaz *et al.* [214] considered a structural design network of connected masses where the design variable is the addition of a unit mass. They sought for optimal choices of mass positioning for the creation of the BG, its central frequency localization, and the control of its width. Afterwards, Bilal *et al.* [215] also used a genetic algorithm to propose a 2D PnCs made of voids in silicon, and presented optimized structural unit-cells with complex design featuring out-of-plane, in-plane and combined out-of-plane and in-plane elastic wave BGs. They were able to attain a BG enlargement of 122.7% for out-of-plane waves. Shortly after, Dong *et al.* [56] considered material composition using Pb or Au in epoxy. They used topology optimization to engineer binary 2D PnC with enlarged BG and reached a relative bandwidth of 108%. **Section 4.5** of this review discusses in detail the optimization methods based on topology optimizations and genetic algorithm engaged for engineering elastic BG through inverse design. Recently, Jia *et al.* [216] proposed a single-phase 2D PnC made of connected masses to reach 119% for the relative BG width, with both theoretical and experimental demonstrations (**Fig. 4(a)**). They also extended their design into 3D lattice of connected spherical masses in a centered cubic configuration and highlighted an omnidirectional BG of 100% relative bandwidth. **Section 3.4** discusses the 3D metamaterial designs proposed in literature for omnidirectional BG engineering.

Regarding guided waves in plates, the first study on BG width maximization was conducted by Halkjær *et al.* [217] in 2006, using theoretical optimization and experimental validation to reach a BG width of 95% (**Fig. 4(b)**). Then, Acar and Yilmaz [212] applied inertia amplification mechanism for engineering wide BG by designing a plate with connected beams. In another study, instead of pillars on one side of the plate, Assouar and Oudich [218] decorated both sides of the plate with a lattice of pillars to increases the coupling between Lamb waves and the resonances of the pillars. They found that the relative bandwidth of the BG can be increased from 27% for single side pillared plate to 51.4% for the double-side pillared plate. Shortly after, Bilal *et al.* [219,220] added holes in-between the pillars to lower the effective stiffness of the plate and increase the coupling mechanism (**Fig. 4(d)**). By this geometrical configuration, they introduce the mechanism of trampoline resonance leading to BG widening from 19% for simple plate with pillars to 48% for the plate with pillars and holes in-between the pillars. BG enlargement was also realized by combining finite structures with overlapping BGs enabling a wider frequency range than what is allowed by individual unit cells [221,222]. Zhu *et al.* [221] showed that combining two different but well-designed resonators in a single unit-cell enables considerable enlargement of LR BG. They designed and fabricated a chiral metamaterial endowed with different resonators to access a BG width of 65% (**Fig. 4(c)**). Celli *et al.* [222] investigated BG widening using pillar resonators fixed on a plate, with different resonance frequencies **Fig. 4(e)**.

Besides, BG enlargement was also realized for SAW, with a first attempt conducted by Rupp *et al.* [223] who applied topology optimization to pattern the surface of a silicon substrate for Rayleigh waves filtering in a wide frequency range. At the geophysical scale, Colombi *et al.* [100] proposed a spatially graded subwavelength resonators distributed on the soil surface, to create what is called a "seismic rainbow" effect. The rainbow effect is based on the coupling between Rayleigh waves and the resonances of stick resonators whose heights are varied gradually to create large LR BGs. This leads to a conversion of Rayleigh waves into bulk waves by deviating the mechanical energy into the earth depth to protect buildings at the surface. A forest with well distributed trees can be used as resonators in the real-world application to achieve the rainbow effect for broad band seismic shielding [100,224].

### 3.4 Three-dimensional structures

The structural design of PnC and metamaterials is manly dictated by the nature of the propagating waves. For the case of guided waves such as Lamb waves and surface acoustic waves, the geometrical construction and material distribution are set to be along 2D space as the wave propagation operates in the plane. Nevertheless, the control of wave propagation along the 3D space is also of great importance, particularly for the case of creating omni-directional BG, known also as complete BG, for the application of vibration shielding. The road to achieve complete BG was enabled by PnC and metamaterials with geometrical and material distributions along the three directions of space.

#### 3.4.1 Composite lattices with embedded units

The first theoretical investigation of band dispersion for elastic waves in a 3D lattice was conducted in 1992 by Sigalas and Economou [6]. They considered a face-centered cubic lattice of solid spheres in an elastic hosting medium (matrix) with contrasting longitudinal and shear wave velocities, by which they demonstrated the existence of a BG. The band dispersion exhibited flat bands at the edges of the BG frequency region which is associated with the local resonance of the spherical units. At that time, this BG was not identified as created by the LR mechanism. Shortly after, the same authors explored the influence of the volumetric filling fraction of the spheres and materials contrast between the spheres and the hosting medium in search for the optimum design of the face-centered cubic lattice for the creation of the BG [80]. Their conclusion held on considering heavy solid spherical units made of either gold or lead and distributed on a Be, or Si or $SiO_2$ matrix with a lattice volume occupation of 10%. While exploring the maximization of the BG, the same group extended their investigations to include simple and body-centered cubic configurations using metallic spheres made of either W, Ni, Fe, Cu, Steel, or Ag inside the epoxy matrix [10]. They observed a relative bandwidth of the BG higher than 50%, with a maximum of 75% by using solid tungsten spheres. This was the first engineering study about BG width maximization in a 3D lattice. Following these works, other studies investigated similar lattices with different materials [81,87,174] or used unit-cells containing both spherical and cubic scatterers [176], or piezoelectric spherical units [177,178]. Further, Suzuki and Yu [82] characterized the dispersion of evanescent waves by developing a numerical model based on the plane wave expansion method to calculate the complex band structure of 3D PnC made of spherical tungsten scatterers embedded in polyethylene matrix.

It was not until the year 2000, that we witness a first experimental fabrication and characterization of a 3D lattice, namely the seminal work of Liu *et al.* [41]. Nevertheless, this classical design of spheres in a hosting media has seen a decrease in interest because embedding heavy spheres in a bulk material is not only challenging and onerous in fabrication with limited choice in materials, but the structure itself can be heavy for practical applications. It was not until 2015, that the field of elastic BG with 3D structural design was revived thanks to the rapid development and wide availability of additive manufacturing techniques.

### 3.4.2 Single phase architected metamaterials

There has been a growing interest in the search for constructing artificial materials that can have more than one functionality, such as combining the low density with mechanical strength. These two properties are highly desirable for aeronautic, aerospace, and automotive applications, mainly for the purpose of material longevity and energy saving. Interestingly, nature provides us with multitude examples of lightweight and strong materials with mind-blowing architectural designs based on highly ordered structuration with hierarchical building from the microscopic scale to the visible macroscopic size. Examples are wood [225], bones, exoskeletons, and spider silk [226]. Drawing inspiration from these biomaterials, researchers have developed and used contemporary fabrication techniques to mimic their micro-structure and access ultra-low materials density with remarkable mechanical behavior. Particularly, additive manufacturing techniques have enabled the fabrication of architected materials with precise geometrical features at multiple orders of hierarchical building [227,228]. Most of these techniques use a single-phase material which means that the constituent material of the architected structure is a single material. This sparked a new generation of artificial materials made of 3D architected lattices with outstanding mechanical properties such as recoverability maximization under compression [229,230], super-elastic tensile behavior [227,228], and even decoupling the density and mechanical performance [231].

Such accurate control over the geometrical features and material composition at the micro-scale in the building process is of great interest from the perspective of wave propagation. 3D printing techniques were employed to create single-phase lightweight architected lattices with outstanding acoustic performance in terms of BG width. Most of these architected designs are of centimeter size lattice constant with low frequency BG (below 10 kHz). Taniker and Yilmaz [117] designed and manufactured a 3D octahedron structure using a polymer printer (**Fig. 4(f)**), that leveraged inertia amplification to enable wide elastic BG. Their lattice constant is around 11cm to reach a very low frequency BG ranging from 70 to 200 Hz. Other works have employed embedded or connected masses to introduce LR BG with increased bandwidth (**Fig. 4(g)-(j), (m)** ), but at the cost of increasing the total mass of the system which may hinder their applications where being lightweight is vital [119–127,129]. For instance, Matlack *et al.* [119] used steel cubic inclusions embedded in a 3D printed cubic frame where the size of a unit cell is 18.25 mm. The structure was fabricated using combined 3D printing and manual assembly by embedding the steel cubes into the printed polycarbonate frame at the middle of the printing process. This meta-structure with heavy masses led to the opening of BGs from 6 kHz to 10 kHz and between 2.15 kHz and 6.11 kHz using different material stiffnesses. In an attempt to reduce the total masse, McGee *et al.* [232] used connected hollow sphere units enabling wide BG for vibration attenuation. More interestingly, a lightweight design using only connected metallic curved beams without heavy masses was proposed by Warmuth *et al.* [233] to enable ultrasound wide BG (**Fig. 3(k)**). The structure was fabricated using elective Electron Beam Melting (SEBM) which is a powder based generative manufacturing technique. Moreover, auxetic designs with added heavy masses have shown to be a promising route for enabling wide BG at low frequency as it was demonstrated by D'Alessandro *et al.* [121] and Fei *et al.* [127] (**Fig. 4(i), (j)**). Very recently, Gerard et *al.* [128] proposed an ultralight auxetic single phase metamaterial made only with thin polymer rods, where trampoline mechanism is extended to the 3D space to create a wide omnidirectional BG (**Fig. 4(l)**). Their structure has a mass density as low as 0.034g/cm$^3$ with a BG width of 82.8% ranging from 2.3 to 5.55 kHz with a lattice constant of 2 cm, which is associated with negative and divergent effective bulk modulus coupled with a near-zero yet positive effective mass density. The fabrication of such auxetic structure was enabled via a high-resolution large area projection micro-stereolithography platform. In another work, Muhammad and Lim [129] proposed a 3D lattice made of connected cylindrical masses on a cubic hollow frame that was fabricated using 3D printing, and reached a BG width of 160% ranging from 1.25 to 11.32 kHz with a lattice constant of 5 cm.

In these architected lattices, not only the precision of the 3D printing process is important, but also knowing the mechanical properties of the constituent material is essential for the desired dynamic behavior of the metamaterial. In fact, despite the advance in additive manufacturing techniques involving a wide range of materials [229,234], polymers have been predominately used in building architected lattices endowed with elastic BG, though polymers also display a wide range of mechanical properties that present some challenges to control. For instance, the stiffness of polymer is dependent on the printing orientation as well as ambient conditions such as temperature. The polymer's Young's modulus and loss factor are also frequency-dependent. These factors have hindered a precise numerical prediction of the dynamic behavior of the printed meta-structure. On the other hand, the advance in additive manufacturing techniques have also enabled 3D printing based on non-polymers such as metals and ceramics [229,234], and even multi-materials printing [235]. These techniques can be leveraged to make elastic metamaterials that are potentially more predictable and with less dissipation, but have been largely unexplored.

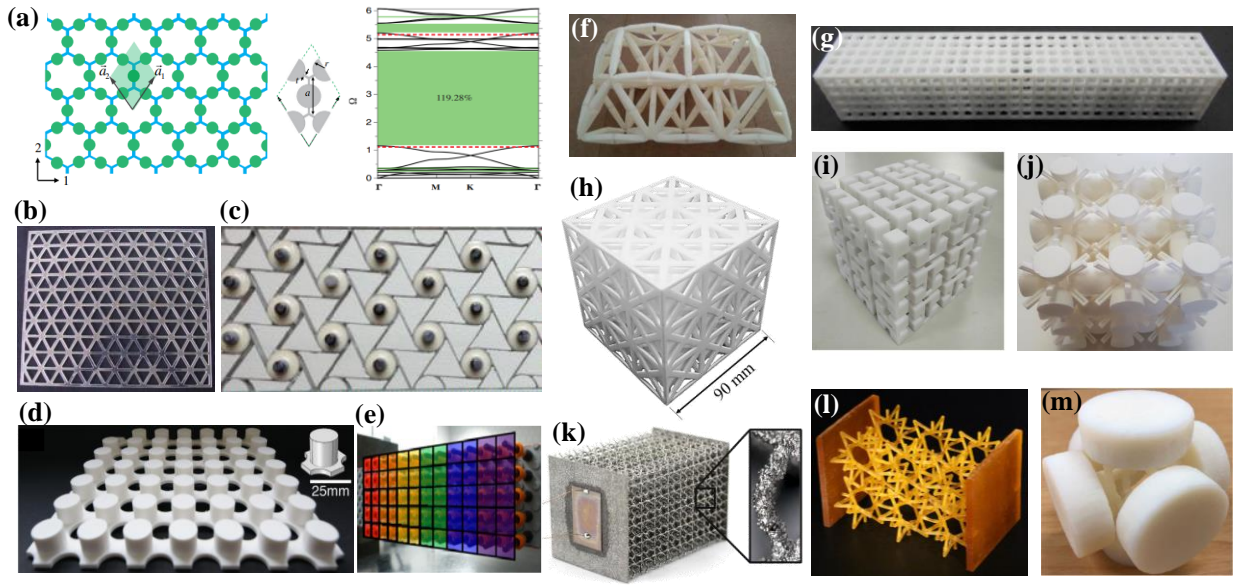

**Fig. 4**. **2D and 3D elastic architected metamaterials.** (a) A phononic lattice made of connected masses for wide elastic BG [216]. (b) An architected phononic plate with triangulate lattice of triangular holes [217]. (c) A chiral elastic metamaterial with resonant units [221]. (d) A trampoline metamaterial plate [220]. (e) A metamaterial plate decorated with low frequency resonators [222] .(f) A 3D metamaterial with induced inertial amplification-based vibration [117]. (g) A 1D metamaterial with integrated heavy mass units [119]. (h) A 3D mass based resonant metamaterial [125]. (i) A 3D anti-chiral auxetic metamaterial [127]. (j) A 3D auxetic metamaterial with connected masses [121]. (k) A 3D metallic metamaterial [233]. (l) A 3D lightweight auxetic metamaterial [128]. (m) A 3D resonating metamaterial [129].

## 4. Contemporary avenues for elastic bandgap engineering

The early works on elastic BGs laid the foundation for understanding the underlying mechanisms of BG and how they are associated with structural design of PnC and metamaterials. This enabled efficient mechanical wave control and have opened avenues exploring new routes for wave physics and engineering elastic metamaterials that aimed at new real-world applications. In this section, we review contemporary efforts where elastic BG engineering is at the heart of the structure dynamic performance. This includes the development of advanced optomechanical systems, elastic topological insulators for robust waveguiding and confinement, acoustic energy harvesters, PnC for sensing, active PnC and metamaterials, and inverse designs for BG engineering.

### 4.1 Optomechanical crystals

For more than half a century, the field of cavity optomechanics has flourished thanks to the plethora of applications in fundamental and applied physics [147,148,151] that hinge on the interaction between electromagnetic radiation and nano or micromechanical vibration. Phononic BG engineering is indispensable in the context of optomechanical device design and has enabled elastodynamic performance by increasing the phonon lifetime and enhancing optomechanical interactions. After the introduction of the concept of BG for classical waves at the end of 80s, we have witnessed an exponential rise in interests toward both photonic and phononic crystals. A combination of the two crystals was born, coined phoxonic crystal (PxC) (the "x" stands for both "t" and "n") which is a dual photonic and phononic crystal that exhibits simultaneously BGs for optical and acoustic or elastic waves [130,131,134–136,139]. Maldovan and Thomas [130] first introduced PxC crystals made of 2D lattices of square and triangular periodicities of either solid cylinders in air or air holes in solid medium. For specific filling fractions, they demonstrated the presence of simultaneous BG for electromagnetic and elastic waves at the same scale of wavelength. This property attracted more attention when Eichenfield *et al.* [132,133] conducted a series of experiments where they designed and fabricated a one-dimensional micro-optomechanical crystal, with a periodicity of 362 nm, endowed with a well-tailored cavity for strong co-localization of 200 THz photons and 2 GHz phonons. Their optomechanical crystal has constituted a powerful photonic and phononic platform for enhanced acousto-optical interaction providing an original sensitive optical measurement of mechanical vibrations [132,133,137,138,140–142,144,146]. The crystal is endowed with simultaneous photonic and phononic BGs with a well-tailored structural defect to host optical and elastic cavity modes inside their associated BGs, which produce highly confined electromagnetic and mechanical energies inside the defect for enhanced optomechanical interaction. A typical example of the device is presented in **Fig. 5(a)** which is made of a 1D PxC nano-beam with a tapered cavity constructed by gradually varying the hole diameters to enable high quality factor optical mode trapping. 2D PxCs with a periodicities from 400 to 600 nm were also introduced where mechanical BGs can be localized from 4 to 10 GHz to achieve high performance phononic and photonic cavities for strong optical excitation and measurement of localized phononic modes **(Figs. 5(b))** [137,138,140,146]. These nano-optomechanical systems were fabricated from silicon wafers using cleanroom nanofabrication techniques where the lattices patterns were realized via electron beam lithography. At the same time, Fuhrmann *et al.* [236] experimentally characterized the periodic modulation of the optical mode's wavelength by a coherent acoustic phonons formed by SAW. Also Gavartin *et al.* [237] experimentally observed the optomechanical interaction in a defect cavity designed in two-dimensional suspended on-ship PxC (**Fig. 5(c)**). Following these works, theoretical investigations were conducted to study the physical mechanism of optomechanical interactions and engineer the phononic and photonic BGs along with the design of the cavity defect in order to produce highly confined elastic and optical modes inside the BG with enhanced optomechanical interaction [143,145,149,150,238,239]. Further, in high precision experimental optomechanics, a phononic shield is designed using a well-tailored

PnC with wide BG to suppress mechanical losses created from coupling between the modes due to symmetry breaking caused by fabrication imperfections at the nano-scale [140,142,144,146,240] (**Figs. 5(a)**). In addition, Fang *et al.* [240] used a nano-PnC (lattice constant of 480 nm) to design waveguides that wire two local optomechanical micro-cavities which allowed for direct phonon exchange (at a frequency around 6 GHz) between the cavities without dissipation (**Fig. 5(d)**). This cavity-optomechanical circuits could be potentially used for performing controlled coherent signal processing. More recently, inspired by the increased importance of optomechanical resonators in the context of sensing [241] and detection of single biological species [242] and molecules [243,244], Navarro-Urrio *et al.* [245] proposed a PxC for the detection of silica sub-micrometer particles with the capability of determining the position of these particles by leveraging a family of mechanical modes known as pinch modes. Another application for optomechanical crystals is the realization of quantum entanglement of mechanical states (**Fig. 5(e)**) [246–248] which could open a new route towards the development of quantum networks based on silicon optomechanical crystals. However, these optomechanical phenomena were observed under cryogenic environment, and has yet to be realized in room temperature for more practical applications.

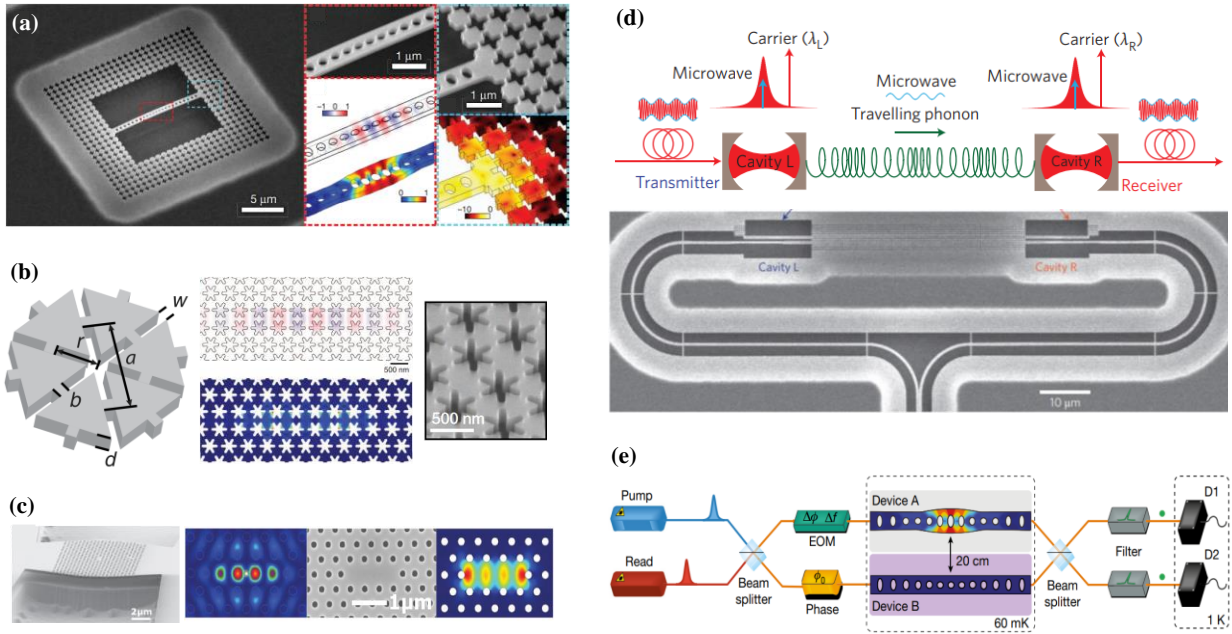

**Fig. 5. Optomechanical crystals.** (a) A phoxonic nanobeam with tapered optomechanical cavity [142]. (b) A 2D PxC with tapered waveguide based on snowflake PnCs [146]. (c) Optomechanical defect cavity [237]. (d) Routing two connected optomechanical cavities for phonon exchange (optomechanical circuit) [240]. (e) Phonon entanglement between two optomechanical nanobeams [246].

### 4.2 Topological elastodynamics

The discovery of topological phase transitions and topological phases of matter by Thouless, Haldane, and Kosterlitz (Nobel Prize in 2016) has become a catalyst for the development of a new class of quantum materials known as topological materials (e.g., topological insulators). These materials, interestingly, display symmetry-dependent topological phase transitions and non-trivial topological characteristics such as non-zero Chern numbers. Over the past few years, photonic and phononic crystals [4,249–251] have become a fertile playground for the exploration of the frontier of topological physics owing to their capability of mirroring some of the quantum-mechanical properties of condensed matter systems. Meanwhile, topological photonic and phononic crystals have opened new gateways for designing devices that can route classical wave energy in a highly unusual and useful way. Topological BGs are the cornerstone of topological physics, and they could arise based on various types of mechanisms, such as the Quantum Hall Effect (QHE), Quantum spin Hall Effect (QSHE), and Quantum Valley Hall effect (QVHE).

The transition of the topological phase is often resulted from a certain symmetry-breaking. For example, as predicted by the Haldane model of graphene [252], breaking the time-reversal symmetry (TRS) will open a topological BG, giving rise to robust one-way edge states that are immune to defects. This phenomenon is associated with the QHE. The topological phase in this model could be characterized by a non-zero Chern number, which is obtained by an integration along the Brillouin zone: $C = \frac{1}{2\pi} \oiint F(\mathbf{k}) ds$, where $F(\mathbf{k})$ is the Berry curvature. Such a topological insulator is known as the topological Chern insulator. Unlike quantum systems, TRS cannot be broken by introducing a local magnetic field in elastic wave systems, and alternative means must be sought. In 2015, Wang *et al.* [253] theoretically studied an elastic topological Chern insulator with Coriolis force. Soon after, a gyroscopic PnC was designed by Wang *et al.* [254], in which the gyroscopic inertial effect breaks the TRS (**Fig. 6(a)**). The experimental realization of the gyroscopic phononic topological insulator was reported by Nash *et al.* [162] in the same year, with the observation of topological edge states (**Fig. 6(b)**). The structure was constructed by considering a honeycomb lattice of magnetically coupled spinning resonators where each gyroscope consists of small dc motor spinning a cylindrical mass and suspended to a plate. Beyond the reported topological edge states, Mitchell *et al.* [255] showed that the tunability of the gyroscopic phononic lattice could bring about complex topology. Besides the periodic lattice, gyroscopic phononic systems even exhibit non-trivial topology in amorphous configurations, producing topological BGs [256]. Since Bragg BG is mostly involved in topological elastic insulators, its frequency range strongly depends on the considered size of the phononic lattice. Early works on demonstrating mechanical topological states used lattices constructed to operate at frequencies ranging from few to hundreds of kilohertz. The objective was mainly to provide an experimental proof of topological manifestations in elastodynamics, which was facilitated by centimeter size structures that can be easily fabricated and characterized [163,164,257–263]. Other designs were later proposed at the micro-scale for the purpose of topological elastic waveguiding at high frequency (from hundreds of megahertz to few gigahertz) for microsystem integration [165,264–267].

In contrast to QHE, the QSHE opens a path to a class of topological insulators without breaking the TRS. This approach hinges on the intrinsic 1/2 spin of the electrons and TRS, which collectively create the Kramers doublet. Elastic wave systems, however, are bosonic that naturally lack the 1/2 spin. In order to overcome this barrier, elastic topological insulators were designed to give rise to pseudospins ±1/2 and artificial Kramers pairs. This class of topological insulators are characterized by a Z2 topological invariant, thus the name Z2 topological insulators [268]. The elastic Z2 topological insulators were first realized by Süsstrunk and Huber using a mechanical oscillator lattice with mode polarizations at very low frequencies (between 2 and 3 Hz) [257]. This mechanism of mode polarizations was soon introduced to multiple elastic platforms such as elastic plates [163,258,259], spring-mass systems [269], and granular media [270], with the observation of helical topological edge states that are immune to certain defects such as cavities and bending. For instance, Miniaci *et al.* [258] designed and fabricated a topological PnC plate made of a lattice of triangular and circular holes producing two Dirac cones at around 100kHz with a lattice constant of 2 cm. This Dirac degeneracy was lifted by breaking the symmetry of the lattice which resulted in the manifestation of spin orbital coupling. They also created a nontrivial interface between two topologically different lattices, which hosts helical edge waves (**Fig. 6(c)**). Another method to design elastic Z2 topological insulator is to leverage modal hybridization [260–264], which forms a double Dirac degeneracy in the band structure. This double Dirac degeneracy can be lifted so that pseudospin states become separated, opening up a topological BG (**Fig. 6(d)**). Zone-folding mechanism provides a simple way to achieve modal hybridization, as two single Dirac degeneracies are folded from the $K$ point to the $\Gamma$ point in the Brillouin zone [271]. This zone-folding mechanism provides a robust route for constructing Z2 topological insulators in elastic plates [260,262–265] and SAW systems [266]. Recently, a 3D elastic meta-crystal that obtained both QHE and QSHE topological phases was explored [272], which showed simultaneously topological surface states and hinge states.

In addition to QHE and QSHE, the QVHE has also been employed to construct elastic topological phononic crystals. The honeycomb lattice of the graphene possesses a Valley degree of freedom, which can be manipulated to break the spatial inversion symmetry (SIS) so that the Dirac degeneracy at *K* point can be lifted to separate two different Valley states. These Valley states are proven to yield opposite 1/2 pseudospins, leading to robust edge states that are immune to certain defects. Local resonance structures have been used to break the SIS in lattices formed by mechanical beams [164,165,273–281] (see example in **Fig. 6(e)**), and soft materials [282,283], where the SIS is broken by the variation of resonators in the different valleys. The valley eigenstates have been observed with opposites 1/2 spins, which are characterized by Valley-Chern numbers. These valley states could be arranged along the interface between two different domains and consequently give rise to topologically non-trivial edge states. In addition to the honeycomb lattice, Kagome lattices have also been shown to host topological valley edge states [284,285] by altering the coupling spring between the masses, which gives rise to topological Stoneley waves. In addition, topological valley phases have been found in hierarchical lattices in a recent work by Han *et al.* [286]. Another feasible method to create elastic topological Valley lattices is by tuning the strain field in materials, which is especially effective in truss-like lattices [287,288]. Such a method has the advantage of high tunability for elastic topological valley insulators. In addition to bulk elastic waves, Valley Hall effect has been realized in surface acoustic wave (SAW) systems [267,289], where pillars were placed in a manner of breaking the SIS of a C6 symmetry lattice. The topological valley phases can be also embedded in systems with other topological phases. For example, Qian *et al.* discovered a new topological regime identified as the Valley-Chern effect by connecting the QVHE system to the QSHE system [290]. Such an elastic lattice was constructed by magnetic spinners with the ability to engineer the Berry curvatures near the valleys, which provided stronger topological protection compared to the original valley Hall designs. In another instance, Mei *et al.* successfully demonstrated a topological beam splitter by an elastic lattice with topological edge states protected by both QSHE and QVHE topological phases [291].

In classical wave systems, non-trivial topological phenomena are usually featured by a reduced dimensional response. For instance, 1D edge states can be engineered in 2D lattices. Non-trivial topological phases in 2D materials bring not only bulk-edge (2D-1D) correspondence but also bulk-corner (2D-0D) correspondence, which could be manifested by zero-dimensional topological corner states. Such a phenomenon often requires a higher-order topology [292]. One possible way to realize the higher-order topology is by introducing a π gauge flux to each plaquette in a square lattice. In 2018, this concept was demonstrated by Serra-Garcia *et al.* [293] via an elegantly designed hopping mechanism in a perturbative mechanical metamaterial. An alternative way is to establish a breathing hexagonal lattice that hosts a second-order topological BG. Fan *et al.* [166] demonstrated a hexagonal high-order elastic topological insulator via beam-coupled masses, followed by another experimental demonstration in elastic plates by Chen *et al.* [294]. High order topology has also been demonstrated via a square lattice in elastic plates [295] (**Fig. 6(f)**). It has also been shown that elastic Kagome lattices can host second-order topological corners states as proved by Wu *et al.* [296] and Wang *et al.* [297]. Additionally, pumping of topological elastic edge states was demonstrated in a plate by modulating its effective stiffness via the thickness where the edge mode in the BG was pumped from one edge to another [298–300].

One of the advantages of exploiting bands topology for the realization of guided elastic waves along edges or interfaces is the immunity to structural defects. Several works investigated the sensitivity of the topological elastic states to structural defects by either creating point, line or arbitrary defects via the removal of several unit cells, or even deforming the lattice [162,163,165,254,255,260,261,265,266]. This ability of defect immunity pushed further for exploring mechanical edge states in amorphous insulators that includes hyperuniform structures, quasicrystals and even random distribution of lattice units.

Besides, topological elastic lattices have contributed to the exploration of Weyl physics as well. While Weyl semimetals are also characterized by lattice topology, they are featured not by a topological BG but rather a doubly degenerate linear band crossing in 3D momentum spaces. As such, the development of this field is beyond the scope of this review paper. The readers can refer to the following papers for more details [301–303].

Finally, exploring and leveraging the topological features of lattices to seek exotic wave manipulation is continuously thriving where BG is at the core of many topological phenomena. In elastodynamics, among multiples routes to be explored are the introduction of topological defects such as disclinations and dislocations for wave confinement and transport, and lattices endowed with engineered long range hopping between unit cells. However, there are still design challenges that need to be addressed such as to introduce chiral symmetry or long range inter-cells interactions.

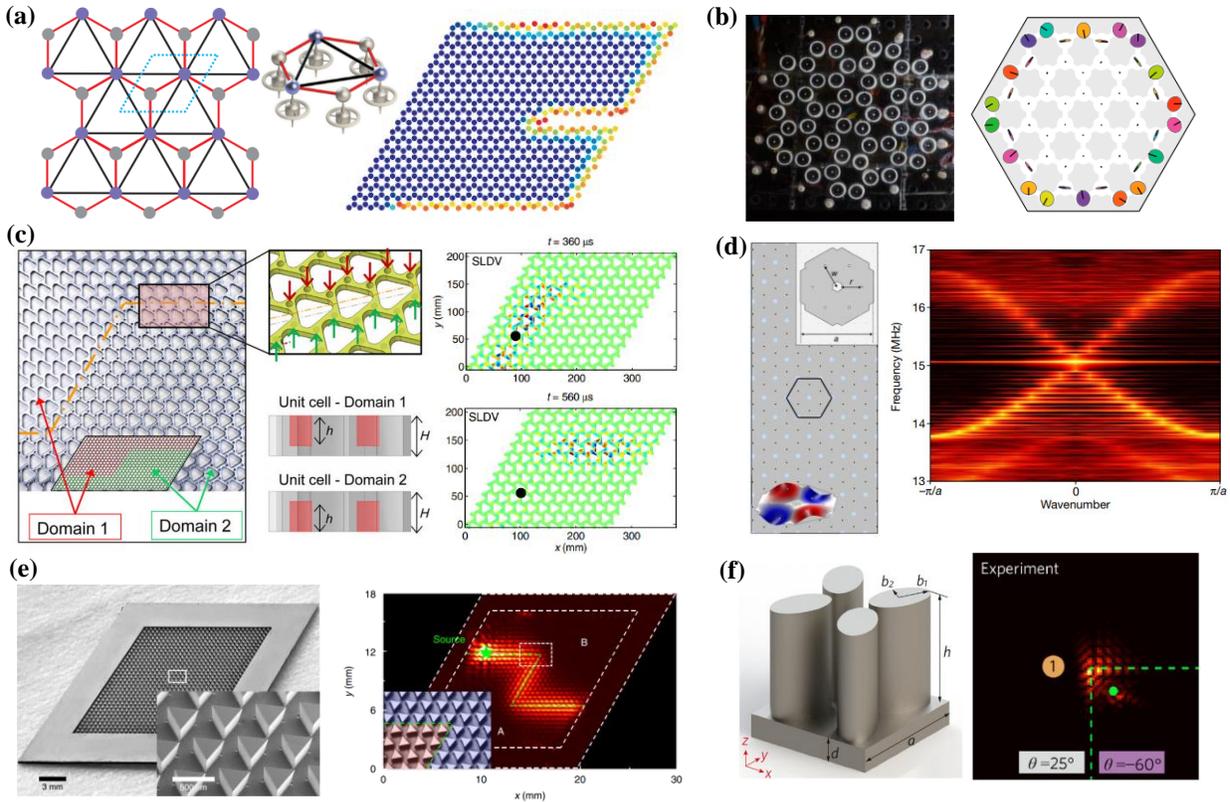

**Fig. 6**. **Topological elastic lattices.** Elastic non-trivial topological lattices (a) Left: Hexagonal lattice constructed by gyroscopes to break TRS. Right: Simulation results of the robust edge state propagation in the latticed shown left [254]. (b) Left: Experimental realization of the gyroscopic lattice. Right: Observed edge states in the gyroscopic lattice [162]. (c) Left: Photograph of a QSHE elastic topological insulator with a zigzag interface. Mid: The domain wall is highlighted in the top panel while the cross-sections of two domains are shown in the bottom panel. Right: Propagation of the topological edge states along the interface of the QSHE elastic topological insulator [258]. (d) Left: Topological 2D elastic nanoelectromechanical lattice. The top inset shows the unit cell, and the bottom inset shows an example flexural mode. Right: Experimental results of the dispersion curves along the topological waveguide [264]. (e) Left: SEM photo of the QVHE topological insulator on a silicon chip. Right: Experimentally measured intensity of the edge states propagation via a zigzag interface [165]. (f) Left: The unit cell of an elastic high order topological insulator in a rectangular lattice. Right: Experimentally measured displacement field of the topological corner mode in the high-order topological insulator [295].

### 4.3 Acoustic Energy harvesting

The motivation to meet the world's need for sustainable and renewable energy has also sparked the interest for acoustic energy harvesting (AEH). The common objective of AEH is to collect and convert the energy carried by sound and vibration to electrical energy, that otherwise would be wasted through conversion to heat. The classical approach for achieving AEH is to use a resonant cavity with a vibrating elastic element equipped with a piezoelectric material to convert the dynamic strain energy into electrical power. The earliest propositions for acoustic energy harvesters used Helmholtz resonators with a vibrating membrane attached to a piezoelectric ring [304], or a cantilever [305], or a series of piezoelectric plates in a straight acoustic tube [306,307]. With the aforementioned development of PnC and acoustic metamaterials for efficient wave manipulation, these concepts also came to light in the context of energy harvesting. PnCs were used to achieve high wave energy confinement through the design of a cavity inside the structure. An acoustic or elastic BG can be engineered to be located at the frequency of interest and a well-tailored structural defect in the geometry can give rise to a cavity mode inside the BG that has a highly confined elastic energy. This energy can be then collected via the traditional utilization of piezoelectrical elements. Wu *et al.* [152] were the first to explore PnCs for AEH. They used a phononic lattice of cylindrical rods with a lattice constant of 4.9 cm, and a unit-cell defect was created by removing a single rod which created an acoustic cavity mode at 4.02 kHz inside the BG. A flexible piezoelectric PVDF film was inserted inside the cavity to harvest the acoustic energy and a maximum power output of approximately 35 nW was collected for a sound pressure level of 100 dB (**Fig. 7(a)**) [153]. Later, Yang *et al.* [154] placed an electromechanical Helmholtz resonator at the center of a PnC cavity and used their coupling to enhance the harvested acoustic energy and generate electrical power reaching 429 µW for a sound pressure level of 110 dB (**Fig. 7(b)**). Nevertheless, these works used Bragg BG so the size of the PnC lattice is relatively large since it is governed by the operating wavelength ($17 \times 17$cm$^2$ with rods length of 10 cm at 5.5kHz [154]). These proposed designs thus still remain cumbersome, and their wavelength-size dependency hinders their miniaturization for integrated AEH devices operating at the audible frequency regime. To overcome this limitation, metamaterials have now become more suitable candidates for AEH due to their deeply subwavelength features. Ma *et al.* [308] proposed a membrane-type metamaterial made of a vibrating membrane with a back cavity, where the acoustic energy around 150 Hz can be absorbed using a resonance state at deep-subwavelength range (device thickness is only of 17mm while the wavelength is around 2.37m in air). The mechanical to electrical conversion is realized by magnet wires carried by the membrane and four pairs of neodymium magnets along the magnet wires to reach an acoustic–electric energy conversion efficiency of 23% (**Fig. 7(c)**). For harvesting the elastic energy from the vibration of a plate, Carrara *et al.* [309] presented a mechanical energy harvester based on elastic metamaterials made of well distributed resonating pillars. They proposed three structural designs with different energy harvesting strategies: pillars with parabolic distribution for wave focusing to realized AEH at the focal point, an acoustic cavity created from a structural defect introduced inside a square lattice of pillars, and a waveguide constructed by removing a line of pillars in the lattice for broadband energy harvesting (**Fig. 7(d)**). The periodicity of the lattice of pillars is 1cm for an operating frequency of 35kHz. Later, Li *et al.* [310] proposed a metamaterial with the dual functionality of sound absorption with over 20 dB of the sound transmission loss and acoustic energy harvesting with a maximum energy conversion efficiency of 15.3% (**Fig. 7(e)**). At the same time, Qi *et al.* [155] proposed a planar AM made of periodic distribution of low frequency pillar resonators over an aluminum plate (**Fig. 7(f)**) and theoretically demonstrated the possibility of AEH at the subwavelength scale. They designed a subwavelength cavity with a defect mode located at 2.26 kHz using a lattice constant of only 1 cm. Considering the same system, Zhang *et al.* [157] and Ma *et al.* [311] presented an experimental demonstration where they reached a maximum harvested power of several microwatts from 100 dB of sound (**Fig. 7(g)**). In these pillared plates, AEH was carried out through a structural defect by removing several pillars to create a cavity mode inside the BG. The existence of the cavity mode is conditioned by the spacing as a result of removing the pillars, and this spacing has to be comparable to half of the wavelength of the flexural wave at the desire frequency. Consequently, a lower frequency of interest implies that more pillars need to be removed in order to create the cavity mode inside the BG which increases the size of the planar metamaterial. To overcome this constraint, Oudich and Li [156] proposed an alternative solution of changing the mechanical properties of the resonators rather than removing them and theoretically showed energy harvesting at deep-subwavelength scale (**Fig. 7(h)**). A defect mode was created in a centimeter size cavity inside the BG at 600 Hz. Though not based of BG engineering, some other periodic structure designs used coiled acoustic channels [312,313], coupled acoustic resonating membranes [314], or coupled defects inside an elastic PnC [315] for the purpose of AEH or elastic energy harvesting. Recently, Javadi *et al.* [316] constructed a PnC consisting of five steel slabs standing in air, with a middle scatterer made of triboelectric nanogenerator, and demonstrated enhanced AEH from the wave confinement caused by the cavity inside the BG. Further, with the advent of PnC that mimic topological insulators, it became possible to create robust elastodynamic edge and interface states inside the BG that are immune to defects, which can be leveraged for acoustic and vibration energy harvesting. Fan *et al.* [317] designed a topological interface state inside the BG created from two sonic crystals having different bands topologies, and used a piezoelectric cantilever for acoustic to electromechanical energy conversion. Very recently, Wen *et al.* [318] designed a phononic plate that harbors a zero-dimensional cavity with a Kekulé distorted topological elastic vortices that were used to realize robust mechanical energy harvesting with a collected electrical power of almost 5 mW (**Fig. 7(i)**). For additional details on this topic, the reader is referred to these previously published review papers [158–160].

Finally, although harvesting ambient mechanical waves is a fascinating idea, these proposed mechanical metamaterials still do not provide sufficient energy for powering large devices. In addition, the range of power shown may be sufficient for low-power devices such as micro-sensors but great efforts are to be made in the structural design of metamaterials to solve the challenge of miniaturization and on-chip integration.

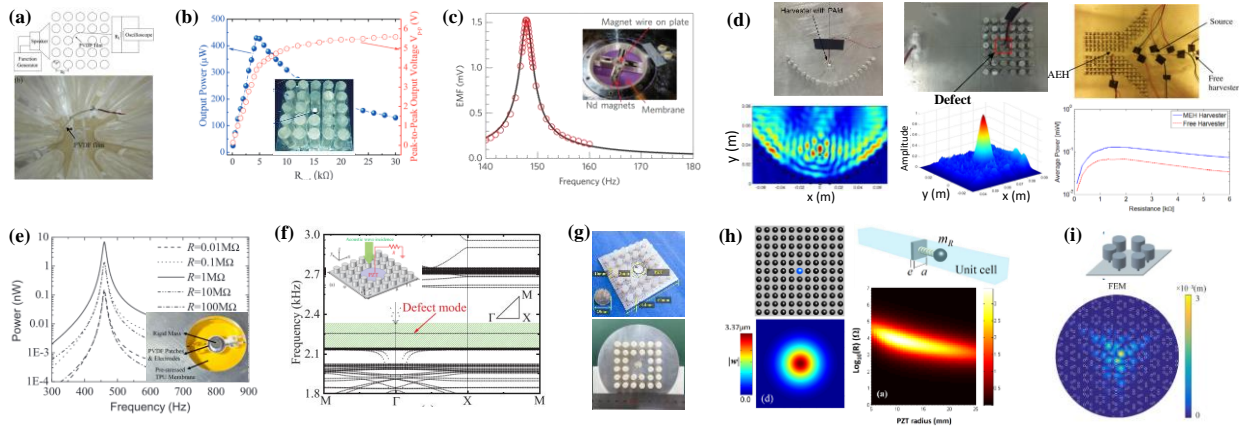

**Fig. 7**. **PnC and elastic metamaterials for acoustic energy harvesting.** (a) A 2D PnC with a cavity where PBDF curved piezoelectric membranes were used to collect the confined acoustic energy [153]. (b) A 2D PnC with a defect made by removing a rod, combined with a HR to enhance the AEH [154]. (c) A membrane-type metamaterial where the mechanical to electrical conversion is realized by magnet wires carried by the membrane and four pairs of neodymium magnets [308]. (d) A pillared PnC with point and line defects for vibration energy harvesting [309]. (e) A hybrid resonant metamaterial for AEH and absorption [310]. (f) A pillared planar metamaterial with a defect created by removing a finite number of pillars, and the band structure showing a defect state inside the BG [155]. (g) Experimental realization of the pillared planar metamaterial [157,311]. (h) A planar elastic metamaterial for AEH with defect resonators [156]. (i) Robust topological cavity in a plate metamaterial decorated with low frequency pillar resonators [318].

### 4.4 Phononic crystals for sensing

Acoustic sensing is also of great interest to the materials science community and could benefit various applications such as medical diagnosis, biosensing, food processing, and underwater detection. Several studies have explored the possibility of using phononic BG for sensing. The sensing mechanism is based on the utilization of localized cavity modes trapped inside a structural defect at frequencies inside the BG, which can be highly sensitive to acoustic and elastic perturbations. Early studies on sensor designs based on PnC utilized lattice constants and cavity sizes in the order of few centimeters for sensing the properties of liquids at operating frequencies raging from hundreds of kHz to few MHz. Lucklum and Li [319] presented a sensing proof of concept using a 1D PnC made of solid plates and liquid filled cavities, and a 2D PnC made of water holes inside an elastic matrix made of aluminum or tungsten. The PnC hosts a liquid filled cavity to analyze the variation of the concentration of 2-propanol in water, which is directly related to the change of the wave velocity in the fluid mixture. The change in the concentration is detected through the variation of the resonance frequency of the defect modes as a function of the molar ratio of propanol on water. The same PnC structure was utilized for the determination of the octane number of gasoline [320,321] (**Fig. 8(a)**). Salman *et al.* [322,323] conducted a numerical study on the determination of ethanol concentration via a linear waveguide made of a line of holes filled with liquid inside a PnC (**Fig. 8(b)**). Amoudache *et al.* [324] used similar design for the detection of the molar ratio using simultaneous photonic and phononic cavity modes inside their associated BGs. Recently, PnC were designed at the microscale and integrated into SAW microsensors to enhance their sensitivity. Bonhomme *et al.* [325] designed a lattice of phononic micro-pillars (6 μm period) with BG tailored to endow the pillars with highly confined elastic modes, and investigated theoretically their potential for the detection of nano-particles in Love wave-bases platform at 250 MHz (**Fig. 8(c)**). They also fabricated a Love-wave based sensor hosting a square lattice of SU-8 micro-pillars (with 9 μm period) where their resonance at 34.47 MHz creates a very narrow LR BG that are highly sensitive to perturbations [326] (**Fig. 8(d)**). Their device was utilized for sensing temperature, mass load induced by micro-droplets, sugar concentration, and the detection of micro-beads. Besides, Sadeghi *et al.* [327] designed and fabricated a clamped PnC membranes at the microscale made of silicon nitride with a structural cavity for thermal sensing. Using laser-based heating, they also observed a frequency shift of both the defect mode and the BG around 600 kHz using a lattice constant of 30 μm. They even observed the frequency tuning of the defect mode at a point where it leaves the BG. Pennec *et al.* [328] used a PnC made of hollow pillars where the interior can be filled with liquid to detect its properties using whispering gallery modes inside the BG (**Fig. 8(e)**). Lately, PnCs made of a lattice of holes in a thick plate were used as a sensing platform by utilizing a honeycomb lattice with a liquid-filled single hole point defect, and a square lattice of liquid-filled holes with narrow BGs [329,330]. The concentration of the ethanol was characterized via either the shift of the defect mode inside the BG using the first phononic lattice, or the narrow BG shift in the second lattice (**Fig. 8(f)**).

The proposed sensing devices based on PnC with integrated cavity are at an early stage and still far from competing with optical sensing devices. However, it was recently shown that the integration of metamaterials such as an array of resonant micro-pillars at the micro-scale in a SAW device could enhance the sensitivity to temperature detection [326] beyond the limit allowed by the classical SAW devices. This could further open routes towards new generation on-chip SAW devices with integrated elastic metamaterials to push the boundary of acoustic sensing down to detecting ultralow-molecular-weight [325], which has only been achieved by plasmonic metamaterials [331,332] and optomechanical devices [243].

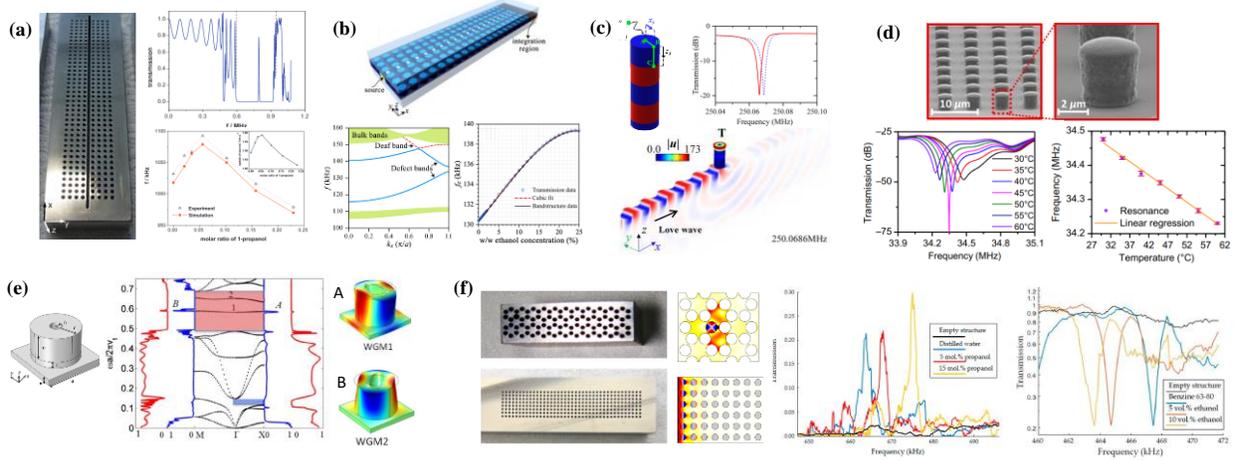

**Fig. 8. PnC and metamaterials for sensing**. (a) (Left panel) A PnC made of holes in a thick slab with a slit defect filled with a mixture of water and propanol. (Right panels) Transmission coefficient showing a BG in which a cavity mode is manifested as a sharp peak. The cavity mode frequency is shown as a function of the molar ratio of the 1-propanol in water [320]. (b) (Upper panel) A PnC with a line defect of small holes filled with fluid which produces defect bands inside the BG [322] (Bottom left panel). (Right panel) Variation of the cavity mode frequency as a function of the ethanol concentration. (c) A phononic pillar made of alternating layers of tungsten and silica producing a BG that hosts edge modes that can be exited with Love wave [325]. (Upper right panel) Transmission spectrum showing sharp dips associated with the edge modes of the pillar. (Bottom right panel) Frequency shift induced by a femto-mass perturbation as function of its position at the top of the pillar. (d) A PnC made of square lattice of micro-pillars in which their resonance frequency is sensitive to temperature variation and micro-particles concentration [326]. (e) Hollow pillar in a plate with whispering-gallery-modes inside the BG used for the characterization of fluids filled into the pillar hole [328]. (f) Two PnCs of honeycomb and square lattices of holes in solid thick plates. The hexagonal lattice hosts a single hole filled with liquid while all holes in the square lattice are filled with liquid for the detection of propanol concentrations. The curves present the frequency response of each PnC as function of the ethanol concentration where we can depict the frequency shift of the defect mode inside the BG using the first phononic lattice, or the narrow BG shift for the second lattice [329].

### 4.5 Active materials for elastodynamics

Though the existence of BGs in PnCs and elastic metamaterials are now associated with a wide range of applications, functional limitations persist as the BG width and frequency localization are constrained by the geometrical and intrinsic properties that are permanently fixed once the structure is designed and fabricated for a specific application. Thus, the scientific community in this field is constantly exploring paths to extend the frequency spectrum of the BGs by incorporating active elements on PnCs and metamaterials. The field of active functional materials has thus been rapidly developed to realize exciting new functionalities like selective wave filtering and non-reciprocity by either statically or dynamically manipulating the intrinsic properties of the metamaterial both in space and time.

#### 4.5.1 Temporally modulated PnC

There has been a rising interest in applying temporally modulated materials to break the time reversal symmetry in order to enable non-reciprocal wave behavior and subsequent active metamaterial-based applications. These time-modulated materials are realized by introducing a controlled time variation of their effective mechanical properties such as the stiffness which can be designed to break the wave reciprocity. Reciprocity is a fundamental property of classical waves in any linear time-invariant media, which stipulates that the measured frequency response of any point remains the same when the source and receiver are exchanged in the considered medium. Non-reciprocity is highly beneficial in the context of elastic wave propagation since it enables unconventional wave functionalities such as unidirectional wave propagation and breaking the time-reversal symmetry for the realization of Chern insulators in classical waves. The idea of temporally modulated periodic media was inspired from the early works of Oliner and Hessel [333] in 1959 and Cassedy and Oliner in 1963 [334]. They introduced a rigorous theoretical method to characterize the dispersion of electromagnetic waves in a medium where the electrical permittivity or the refractive index undergoes a spatiotemporal modulation (STM) in the form of a propagating wave. This type of STM process breaks the time reversal symmetry in the lattice and consequently enables a unidirectional BG. However, the technical capability of building an experimental demonstrator at that time hindered the interest for this class of active artificial materials. Later, in 1998, Winn *et al.* [335] theoretically demonstrated that the spatiotemporal variation of the dielectric constant enables optical band transitions analog to electronic ones in metals and semiconductors. It was not until 2008 that Dong *et al.* [336] first experimentally performed direct photonic transitions using ultrafast tuning of the refractive index where the time interval of this tuning was on the order of the inverse of the frequency difference between the optical modes. Shortly after, Yu and Fan [337] introduced an on-chip optical signal isolation that was achieved by the STM of the refractive index. These studies revived the interest towards time-varying mediums with the emergence of multiple studies on creating new photonic platforms with STM materials. Time-varying photonic meta-surfaces then emerged to show exotic optical functionalities such as unidirectional electromagnetic induced transparency [338], strong and broadband nonreciprocal wave transmission [339–341], frequency mixing [342], compact flat prism [343], and controlled linear frequency conversion [344]. Drawing inspiration from this progress in photonics, a flurry of spatiotemporal phononic designs were proposed in acoustics and elastodynamics. Fleury *et al.* [345] proposed an acoustic non-reciprocal isolator made of three ports system with a central ring cavity endowed with an internal circulating fluid flow to introduce an acoustic bias. Shortly after, they presented a compact acoustic circulator based on space-time modulation of the effective acoustic refractive index [346]. However, these pioneer acoustic systems do not involve the creation of BG based on phononic lattices. An acoustic analog of STM medium of Cassedy and Oliner [334] was proposed in elastodynamics by Trainiti and Ruzzene [347] who theoretically investigated the dispersion of elastic waves propagating in a beam with STM materials properties, for both longitudinal and bending waves. By spatiotemporally modulating the Young's modulus of the beam in a waveform shape, they demonstrated a new class of unidirectional BGs. In fact, starting from a simple PnC with a specific BG designed from a spatial modulation of the Young modulus, the introduced time variation shifts the BG to higher frequencies for wave vector directed along the propagating modulation while the same BG shifts to lower frequencies for wave vector in the opposite direction. This study sparked a host of theoretical investigations to explore the physics of non-reciprocal BGs in acoustic time-varying structures [348–351], while other works took the challenge of experimentally engineering mechanical materials with time and space modulated effective properties. These active materials have unit cell sizes in the order of few centimeters leading to operating frequencies ranging from 10 Hz to 20 kHz, since the objective was to mainly demonstrate non-reciprocal wave propagation. Wang *et al.* [352] were the first to experimentally achieve such STM using a 1D periodic coupled permanent magnets (period of 33.4 mm) and coils where the coupling can be varied in time via alternating current (AC) in the coils which dynamically modulates the magnetic force between these coils and the magnets (**Fig. 9(a)**). The system here, could be modeled by a series of coupled masses connected to the ground by springs with time modulated effective stiffness. This magnet-coil based approach was proven to be effective to dynamically modulate the effective stiffness with high speed which led to non-reciprocal dispersion of longitudinal elastic waves observed around 20Hz. Using the same technique, the same group proposed an elastic metamaterial plate made of periodic modulated cantilever resonators distributed on a beam. Each resonator is built from a permanent magnet and a coil mounted on a flexible cantilever unit fixed on the beam, which lead to a demonstration of non-reciprocal propagation of flexural waves (**Fig. 9(b)**) [353]. Another interesting modulating approach for the effective stiffness was proposed by Ruzzene *et al.*, who showed unidirectional BG for flexural waves using piezoelectric elements periodically distributed on a beam and connected to controllable electrical circuit (**Fig. 9(c)**) [354–356]. The experimental realization showed unidirectional wave propagation at 10.5 kHz with a periodicity of 24 mm [356]. The same approach was also investigated in depth by Yi *et al.* [357]. Meanwhile, an interesting yet different STM approach was introduced by Wallen and Haberman [358,359] who used nonlinear large mechanical deformations to spatiotemporally vary the effective stiffness in a periodic structure to enable non-reciprocity wave dispersion. Also, Wang *et al.* [360] considered a prestressed periodic structure made of prismatic tensegrity cells, and they modulated the pre-stress both in space and time to break the time reversal symmetry and achieve unidirectional BG. Attarzadeh *et al.* [361] used a series of local resonators in which their effective stiffness was modulated by varying the second area moment of inertia of each resonator's arm through dynamically changing its angular orientation (**Fig. 9(d)**). Besides, non-reciprocity is also realized in the case of mechanical analog of Floquet topological insulator (FTI). Darabi *et al.* [362] built a FTI made of a hexagonal array of piezoelectric disks distributed over a plate surface and shunted through electrical circuits controlled to modulate their effective capacitance both in space and time in a way to break the time reversal symmetry and induce topological protected edges states with unidirectional propagation (**Fig. 9(e)**).

Non-reciprocal wave propagation was also demonstrated for SAW without relying on the presence of BG from the STM, but rather on altering the propagation characteristic by applying an electrical field [363], or an external magnetic field [364–366], or by leveraging the magnetoelastic coupling [367].

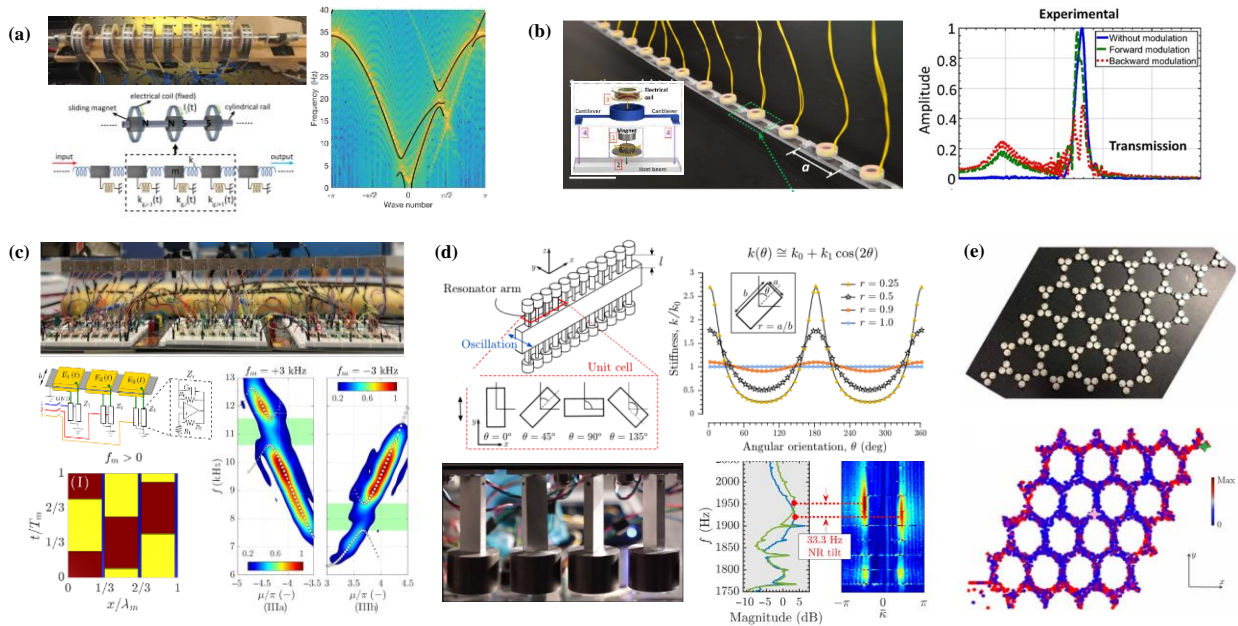

**Fig. 9**. **Non-reciprocal PnC.** (a) A nonreciprocal 1D PnC made of an array of permanent magnets and electromagnets to modulate the effective stiffness in space and time [352]. (b) A 1D metamaterial with periodic resonators made of coupled magnets and coils to modulate the local effective stiffness [353]. (c) A nonreciprocal phononic plate for flexural waves with spatiotemporally modulated effective bending stiffness via shunted piezoelectric periodic elements [356]. (d) A 1D metamaterial beam made of a series of resonators where the effective stiffness is modulated via the change in the second area moment of inertia of each resonator's arm through dynamical rotation [361]. (e) An elastodynamic Floquet topological insulator with one-way edge state propagation enabled via the modulation of the piezoelectric units of the lattice [362].

### 4.5.2 Stimuli-responsive metamaterials

Owing to the rich physics and applications associated with elastic wave BGs, several studies began exploring the possibility of tuning their widths and frequency of occurrence in real-time. Preliminary efforts towards this direction comprised the theoretical demonstration of BG control through varying the unit cell's shape [368–370] or the elastic properties [371–376]. However, enabling such geometrical or material property variation requires the direct or indirect influence of an external force which gave birth to the sub-field of stimuli responsive phononic crystals and metamaterials [377–379].

An early study by Goffaux and Vigneron [380] showed that the Bragg BG for a lattice consisting of solid square-section columns, could be tuned by simply rotating all the columns about their vertical axis. This inspired several subsequent works that theoretically demonstrated the tailoring of both Bragg and resonance-based BGs through simple geometrical modifications like varying the shape and symmetries of scatterers [381,382] or incorporating inclusions with elastic anisotropy [383,384], or adding an additional inclusion [385]. In this line of work, an interesting study by Bertoldi and Boyce [386] demonstrated that periodic elastomeric lattices could serve as a host for tailorable phononic BGs due to their ability to exhibit mechanically triggered pattern transformation. This finding sparked interests for instability/nonlinearity-induced and deformation-dependent BG [387–390] that were later also experimentally realized by Wang *et al.* [369], Shan *et al.* [391] and Babaee *et al.* [392] (for airborne sound). The first experimental study in this regard [369] proposed an adaptive elastic metamaterial whose unit cells comprise resonators that consist of metallic cores connected to easy-to-buckle elastomeric beams (**Fig. 10(a)**). The metamaterial was fabricated using silicon rubber and a mold-casting process where the mold was built via 3D rapid prototyping. The BG was opened at around 100 Hz using a lattice constant of 50 mm. Such a system allows for the unit cell to deform systematically and change shape significantly, upon compression. This in turn tailors the BG and enables deformation dependent BGs in the band structure and transmission measurements. However, these works rely on large contact mechanical forces applied to the sample and hence had limited applications in their current form.

In parallel, other studies probed the alternative strategy of tuning the BGs by varying the intrinsic properties of the material constituting the phononic lattices. Although theoretically similar, this alternate approach requires the presence of an external field-like stimuli such as magnetic field [371,372] that does not deform the lattice and hence could be more promising for real-world scenarios and for other research that is also discussed in the previous sub-section of this manuscript. Early studies in this context, theoretically showed that BGs could be tuned by modifying the lattice to incorporate materials whose stiffnesses could be actively tuned via optical fields [393], electrorheological materials [394], electrical [395], piezoelectrical circuits [373,396–399], external magnetic field [371], and temperature variation [400]. The first experimental validation of such an active metamaterial was put forward in the study by Bergamini *et al.* [401] (**Fig. 10(b)**) who considered a structure comprised of successive cylindrical stubs (1 cm period) with piezoelectric discs that were shunted through an inductive circuit in order to obtain frequency dependent stiffness elements between the substrate and the stubs. This in turn allows for tailoring the effective periodicity of the sample that could allow for the generation of a passband within the otherwise wide-band gap. Likewise, a study by Wang *et al*. [402] showed that similar behavior could be achieved by employing electromagnets in a 2D lattice that would attach or detach in the presence or absence an external magnetic field (**Fig. 10(c)**). This is highly beneficial since it implies that each unit cell can be controlled independently, and the BG can be readily turned on or off. The metamaterial is relatively large with a periodicity of 17 cm for an operating BG switching functionality from 5.5 to 12 Hz. Capitalizing on this convenient mechanism, the authors of this work demonstrated tunable digital metamaterials for low frequency elastic waveguiding and isolation.

Intrigued by these simple yet insightful phenomena demonstrated for active 2D periodic designs, researchers began extending these concepts to more complicated systems to enable new applications. For instance, Cha and Dario [395] proposed a novel nanoelectromechanical flexural phononic crystal (**Fig. 10(d)**), that consists of free-standing nanomembranes with circular clamped boundaries. With a lattice constant of 7 μm, the system exhibits a BG that is shown in to be lowered in frequency from 18 to 14 MHz by means of applying a dc gate voltage that creates voltage dependent onsite potential. Further this study also showed that when a dynamic modulation of the voltage is employed, it triggers non-linear effects which induce the formation of a new BG that is analogous to the Peierls transition in condensed matter [403]. Subsequently, active metamaterials were also scaled upward with the advent of more advanced additive manufacturing techniques. To this end, Pierce *et al.* [404], employed a direct ink write fabrication method to produce a metamaterial whose struts are made up of a magnetoactive elastomer. This allows the sample's intrinsic elastic properties to be controlled via an external magnetic field (**Fig. 10(e)**), thereby enabling a BG shift that is dependent on the strength of the applied magnetic field. Similarly, Gliozzi *et al.* [405] fabricated a polymeric material via a UV polymerization method that employed a Methyl red as an azo-dopant, which served both as an optical absorber as well and active light responsive component of an elastic metamaterial. Exploiting this unique characteristic of the intrinsic material, the study here illustrated that the eigenmodes of the resonant pillars are affected by local illumination such that the overall BG can be significantly widened as function of time (**Fig. 10(f)**). Using parallelepipedic pillars with 5 mm between two successive pillars, the BG located at 80 kHz was enlarged from 16 kHz bandwidth without illumination to 38 kHz with pillar illumination. Another realm of studies involved those that combined the aforementioned mechanisms by employing an external field upon the intrinsic material in order to induce a change in geometry of the metamaterial unit cell. The early studies in this regard were those of Foehr *et al.* [406], that leveraged the concept of spiral phononic plates which possess wide BGs that could be tailored based on their geometrical state. This concept was coupled with an external magnetic field in order to realize programmable phononic metasurfaces that have on and off states through flat and programmed scenarios respectively (**Fig. 10(g)**). The presented design [407] consisted of Archimedean spirals with magnets in the center that enable a field-responsive geometry which results in the field-dependent BG ranging from 90 to 140 kHz for a lattice constant of 1.25 cm. Interestingly, this bi-stable metamaterial was also later extended to demonstrate a transistor like device capable of performing logic gate calculations and cascading elastic vibrations [408]. Such a design is highly desirable for the future of advanced computational systems. Likewise, similar concepts also emerged for higher order elastic metamaterials. Montgomery *et al.* [409] presented a magneto-tunable metamaterial that exhibits a considerable change in metamaterial shape through deformation mode branching. This change illustrates a variety of different BGs for various combinations of magnetic field and deformation. However, this study only characterized the static performance of these geometries and not the elastic wave transmission. A similar study conducted by Xu *et al.* [410], however, fabricated and experimentally characterized the elastic wave propagation through a minuscule 3D metamaterial whose intrinsic material was a magneto-elastomer (**Fig. 10(h)**). The geometry under consideration was a well-known negative stiffness lattice that has configurable shapes for different magnetic fields, resulting in different transmission ratios. Additionally, it was demonstrated that for each of the shapes, the system had a different "ON" state which corresponds to the only frequency that is allowed to propagate within the lattice.

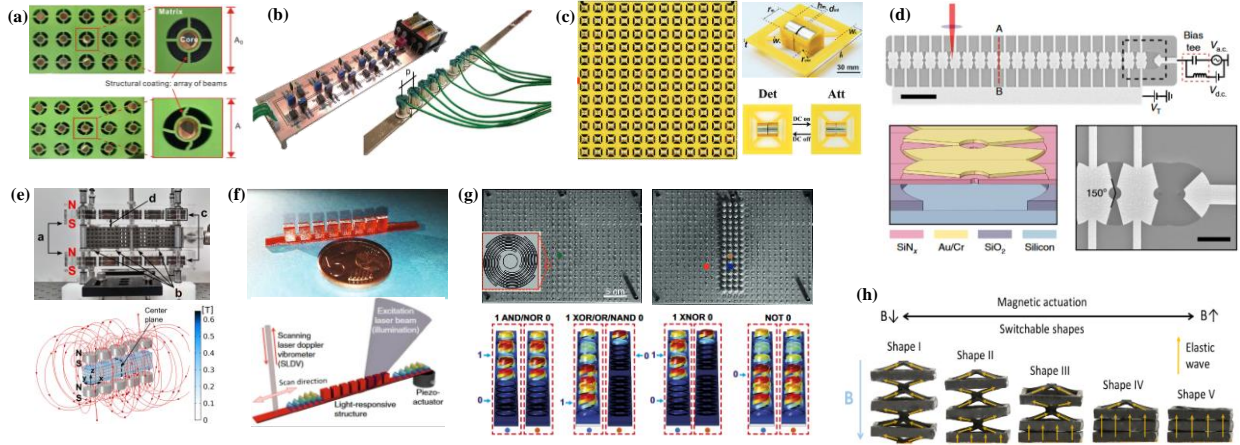

**Fig. 10. Active elastic metamaterials.** (a) A metamaterial made of tunable resonating units. Each unit cell consists of a metallic core connected to a polymeric host via thin elastic beams. The buckling of the beams is exploited to control the elastic wave dispersion and the BG [369]. (b) A 1D PnC made of pillars with piezo-electric discs shunted through an inductive circuit in order to vary their effective stiffness [401]. (c) A 2D active metamaterial where each unit cell is made of two switching electromagnets between the attaching (1 bit) and detaching (0 bit) states to control the BG and the waveguiding functionality [402]. (d) A 1D nano-electromechanical phononic lattice made of periodic free-standing nanomembranes with clamped boundaries. The phononic BG can be shifted by the application of a static voltage [395]. (e) A 1D mechanical metamaterial made of magnetoactive elastomer that reacts to external magnetic field to continuously tune the elastic BG [404]. (f) A 1D metamaterial made of photosensitive parallelepiped pillars where illumination affects the BG [405]. (g) A Programmable PnC made of units of Archimedean spirals with magnets to realize phononic transistor-like device through switching and amplification of elastic vibrations [407,408]. (h) A shape reconfigurable elastic metamaterial through the application of magnetic field to control the attenuation of elastic waves [410].

### 4.6 BG engineering through inverse design

Owing to the fast development of computational capabilities that have emerged over the past decade, several recent studies on PnCs and elastic metamaterials have focused on harnessing new tools for their design and optimization. This has led to inverse-design based studies that start with the desired BG configuration or functionality and employ optimization approaches to arrive at the geometry and material that is required to achieve them. Most of these efforts largely rely on topology optimization [56,57,411–436], genetic algorithms [215,432,437,438], or machine learning based approaches [439–446] and have unveiled unusual and hence previously inconceivable geometries that enable materials with enhanced BG characteristics. While these efforts are now burgeoning thanks to modern computational power, one of the earliest fruitful strides in this direction for elastic waves, can be attributed to the works of Sigmund and Søndergaard Jensen [411] in the early 2000s, who first put forward a theoretical framework showing that phononic BGs could be considerably enlarged by opening the design space of the unit cell geometry, while enforcing the boundary conditions as constraints – in other words, via topology optimization. This has led to a wide variety of subsequent works that later employed topology optimization for both Bragg [418] and local resonance-based BGs [416], and the merging of these efforts for exotic features like directional wave propagation [427,428], negative refraction [420], mode conversion [419] ultra-wide BGs through dielectric elastomers [414] and enhanced inertial amplification [415]. More recent works in this context have also extended the scope of topology optimization to 3D metamaterials [412,447] and by incorporating geometrical uncertainties [424,426,448] in their calculations in order to better capture the real-world scenarios. The calculations in all the above-mentioned studies can be broadly described as coupling of conventional wave analysis methods like multiple scattering theory, finite element, or plane wave expansion with gradient-based or non-gradient-based optimization techniques in order to achieve single or multiple desired objectives [449]. The entire process is thus two-stage - the first is the band structure analysis to extract the dispersion behavior of the periodic crystal, and the second is the mathematical solution of the optimization model that offers a new set of variables a step closer to the desired objective. The first and second stages are thus carried out repeatedly until the system reaches the required criterion. The difference between the two optimization techniques is that gradient descent-based methods [425,450] hinge on a smooth functions which work through continuous variables, while non-gradient-based methods [416,451,452] are often non-smooth and hence runs through somewhat random variables. Non-gradients would thus be more precise and flawless at the cost of higher computational time since gradient-based methods could get stuck at an undesirable point in the local parameter space. Two popular non-gradient optimization techniques are genetic algorithms and simulated annealing, and **Figs. 11(a)**, **(b)** and **(c)**, show representative examples of works that employed these for BG optimization in elastic waves. **Figure 11(a)** shows the work of Jung *et al.* [416], that designed a novel plate-type metamaterial with wide BGs for elastic waves via simulated annealing. The authors here demonstrated the benefit of using simulated annealing over gradient-based topology optimization by illustrating that it prevents the occurrence of realistic design issues and enables BGs at desirable frequency ranges via the unusual unit cell geometries shown here. Similar to this work, Bilal and Hussein [215] also showed that genetic algorithms could serve as a route to realize ultrawide BGs in 2D metamaterials, via the combination of in-plane and out-of-plane waves. This was made possible due to the unique geometrical configurations of the resultant unit cells, as can be seen in **Fig. 11(b)**. Likewise, **Fig. 11(c)** shows the results from a work by Dong *et al.* [437] that carried out a multi-objective optimization of two-dimensional phononic crystals through a genetic algorithm. The two objectives in this work were to realize ultra-wide bandgaps and reduce the mass of the crystal.

Another optimization approach that has been equally popular for the inverse-design of BGs are those that are based on machine learning techniques (ML). These methods focus on acquiring vast collection of training data and utilizing artificial neural networks to classify them and arrive at the required parameters for the desired goal [453]. This route has become increasingly useful due to the high computational power that is now available and its ability to calculate band structures for a large number of geometries. While a majority of such inverse-design based studies are for electromagnetic [442,454–456] and acoustic waves [457–462], some recent works have begun employing ML for mechanical metamaterials and elastic wave BG characterization as well. In the context of mechanical metamaterials with exceptional static properties, ML based inverse design approaches hold great value, since they allow designers to take geometric and material property variation and uncertainty into account in their training data set [463–466]. Inspired by such efforts, several recent studies have emerged centered around employing ML for the inverse-design of elastic metamaterials. It was recently shown in a work by Liu *et al.* [467] that ML could help design PnC with wide BGs, as shown in **Fig. 11(d)**. This was made possible here by taking advantage of neural networks for both the eigenvalue problem (i.e., to calculate the dispersion behavior, as shown in Refs [443,468]) as well as the optimization step. Likewise, a recent study by Wu *et al*. [469] put forward the design of a modular metamaterial shown in **Fig. 11(e)**, which was made possible by combining neural networks for the band structure calculation and genetic algorithms for the optimization, as illustrated. This allows them to realize on-demand wide and tunable BGs. Much more recently machine learning based designs have also found their way for the realization of topological edge states [470] through a recent effort that illustrated that these could be used to judiciously tailor the bandgap for flexural waves in plates.

For additional details on each of the optimization techniques discussed here, the reader is referred to the previously published review papers [413,441,449].

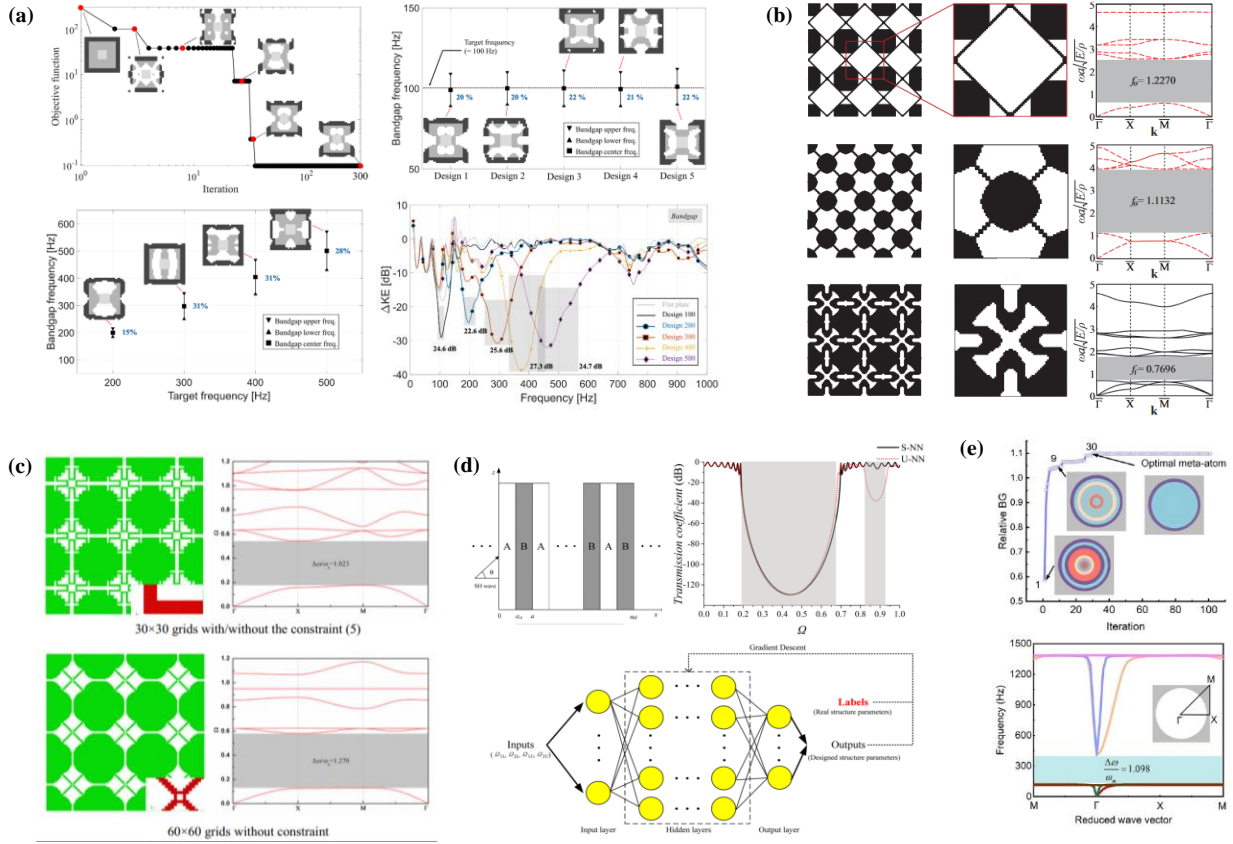

**Fig. 11. Inverse design of elastic metamaterials for BG engineering.** (a) Plate type metamaterials designed by topology optimization [416]. (Left to right, top to bottom panels) Iterations to minimize the objective function; multiple design solutions for BG maximization; Design solutions for five different target frequencies; Kinetic energy differences ΔKE and the BG for 5 designs. (b) 2D PnCs designed using genetic algorithm for BG maximization [215]. (c) 2D PnCs designed with multi-objective optimization with simultaneously maximal BG width and minimal mass [437]. (d) 1D PnCs optimized for BG enlargement using neural network [467]. (e) 2D PnCs with meta-atom optimized for BG enlargement using combining genetic algorithm and machine learning [469].

## 5. Conclusion and perspectives

In conclusion, we have presented an overview of the history, including the latest development of PnCs and elastic metamaterials, focusing on their characteristics derived from frequency bandgaps. This review paper starts from the basic principles of the two types of bandgaps: Bragg and local resonance bandgaps. It then attempts to provide an extensive survey of existing 2D and 3D phononic crystal or metamaterial designs that possess bandgaps. Finally, we discuss the state-of-the-art of bandgap engineering, which encompasses GHz phononic crystals that can interact with photons, bandgaps that are topologically nontrivial, energy harvesting from vibration, sensing, active materials that can be modulated or reconfigured by external fields, and bandgap engineering via inverse algorithms.

Going forward, band gap engineering will likely remain a central topic in phononic crystals and elastic metamaterials, driving the advancement of structured materials for controlling elastic waves. In particular, it is expected that elastic bandgap engineering will continue to benefit from the exciting development of topological physics, which has already led to the realization of elastodynamic Z2 topological insulators, higher-order topological insulators, and Valley Hall edge states, among others. For example, topological defects such as disclination and dislocation can host bound states within the bandgap, that trap energy in the bulk rather than at the corner of the material. A recent paper demonstrated that disclinations in an acoustic lattice can give rise to degenerate zero-energy bound states [471]. It would be interesting to also demonstrate such symmetry-protected bound states in elastic wave systems for sensing, energy harvesting, or photon-phonon interaction. However, an elastodynamic structure that yields chiral symmetry must be first conceived.

Further, elastic BG engineering plays a vital role in the design of optomechanical crystals for mechanical modes trapping and waveguiding at the GHz regime. Thanks to the progress made in the field of nanofabrication, the field of optomechanics has seen a significant development in the last decade with significant discoveries and experimental achievements that have closed the gap between the quantum-mechanical and classical-wave worlds towards building advanced systems that elegantly unite photons and phonons [151]. This includes the realization of non-classical correlations between single photons and phonons [472], optomechanical entanglement [246], quantum transduction [473], optomechanical quantum teleportation [474], reconfigurable quantum phononic circuits [475], among others, which could facilitate the development of future quantum based devices for quantum communication, quantum memories, and quantum transducers. However, most of these recent quantum optomechanical demonstrations were performed in cryogenic environments in order to reduce thermal noise. A challenge to overcome in upcoming years will be to demonstrate the aforementioned quantum phenomena in room-temperature for more practical applications [476].

On the other hand, architected metamaterials embody the promise for next-generation 3D bandgap engineering. While the vast design space of architected metamaterials could enable uncounted designs with intriguing dynamic properties, this advantage can also become a hurdle for traditional forward design approaches. This is further confounded by the fact that these materials' dynamic footprint cannot be adequately described by singular peak values such as transmission rate, but are in fact captured by their entire dynamic behaviors spanning a wide range of frequencies. These challenges, however, create a great opportunity for data-driven approaches to overcome difficulties that make forward design approaches extremely ineffective. 3D architected metamaterials can further benefit from progresses made in the additive manufacturing domain, such as to leverage multi-material printing [235] as well as piezoelectric material printing [477] to further expand the design space of these rationally designed structures.

Furthermore, non-periodic lattices such as quasicrystals and hyperuniform structures can also serve as interesting platforms for controlling elastic waves. Although the lack of periodicity hinders the possibility of precise dispersion characterization via the band structure, these aperiodic lattices can still display frequency ranges where the wave cannot propagate. For instance, using frequency response analysis, it was demonstrated that quasicrystals can display photonic and elastic bandgaps and host intrinsic optical and mechanical wave localizations without introducing any structural defect [478–480]. Also, the field of hyperuniform structure has long attracted significant attention in photonics for creating bandgaps, waveguides and energy localization [481–483]. However, few works have introduced these aperiodic lattices [256,478,479] in elastodynamics, and exploration of their dynamical performance with real experimental demonstration is lacking for potential applications such as optomechanics.

Recent development on bilayer phononic crystals has also pointed out a new direction to engineer the bandgap using interlayer coupling and potentially even the twist degree of freedom. A recent study displayed a twisted bilayer design for elastodynamics made of two lattices of pillars distributed on both sides of a thick plate. Surface acoustic waves can propagate on each side of the plate with coupled wave dispersions between the two phononic lattices [484]. It was demonstrated that at a specific twist angle that makes the bilayer phononic crystal have an even sublattice exchange symmetry (even SE), an elastic bandgap can open. This particular bandgap was shown to host high order topological state (corner modes) in a photonic bilayer system [485]. Furthermore, the interlayer coupling can be tailored to create an elastic bandgap that can host Valley Hall topological interface states at a frequency range unattainable by the monolayer phononic crystal. These findings suggested that bilayer phononic crystals can have extended dispersion capability via the twist and inter-layer coupling, which could open new routes towards a new class of phononic heterostructures for exotic wave phenomena.

Meanwhile, nonlinear waves including solitons could constitute another promising area of research that can benefit from elastic bandgap engineering. Particularly, granular crystals have been extensively explored as they host highly nonlinear waves, which have enabled the design of impact absorbers [486], lenses [487], switches [488], and nondestructive detection [489]. Until now, elastic metamaterials and PnCs have been mostly designed to control linear elastic waves. The ability to extend their functionality for manipulating nonlinear elastic waves is of high interest. Recently, a deformable metamaterial was built to support the propagation of vector solitons with two polarizations which was used to design an architected lattice with bandgaps for solitons [490,491]. These studies open the avenue towards a future generation of architected metamaterials with engineered nonlinearities to achieve amplitude-dependent control of vibration.

Another direction of exploration in elastodynamics is lattices designed in non-Euclidean space. For instance, hyperbolic lattices have been introduced in circuit quantum electrodynamics with the manifestation of exotic phenomena such as flat bands and localized eigenstates [492]. This has sparked the exploration of hyperbolic band theory [493] and crystallography in non-Euclidean geometries, which spanned the lattice space beyond the classical Bravais lattices [494]. Recently, theoretical explorations were conducted for wave dispersion in hyperbolic photonic lattice [495] as well as in elastodynamics [496]. This field presents several exciting avenues of research for exploring wave physics in curved space that is accessible via origami designs for instance, which could expand the design space toward curved elastic metamaterials.

Overall, as the field of bandgap engineering for elastic waves grows and evolves to be more multidisciplinary, we fully expect that this subject will continue to strive for decades to come.

**Acknowledgements**

Y. J. thanks the NSF for supports through CMMI 2119545, 1951221 and 2039463.

**Conflict of Interest**

The authors declare no conflict of interest.

**Data Availability Statement**

Data sharing not applicable to this article as no datasets were generated or analyzed during the current review.

**Keywords**

Phononic crystals, Elastic metamaterial, Bandgap.

[302] X. Shi, R. Chaunsali, F. Li, and J. Yang, *Elastic Weyl Points and Surface Arc States in Three-Dimensional Structures*, Phys. Rev. Appl. **12**, 024058 (2019).
[303] S. S. Ganti, T.-W. Liu, and F. Semperlotti, *Weyl Points and Topological Surface States in a Three-Dimensional Sandwich-Type Elastic Lattice*, New J. Phys. **22**, 083001 (2020).
[304] S. B. Horowitz, M. Sheplak, L. N. Cattafesta, and T. Nishida, *A MEMS Acoustic Energy Harvester*, J. Micromechanics Microengineering **16**, S174 (2006).
[305] S. Noh, H. Lee, and B. Choi, *A Study on the Acoustic Energy Harvesting with Helmholtz Resonator and Piezoelectric Cantilevers*, Int. J. Precis. Eng. Manuf. **14**, 1629 (2013).
[306] B. Li, J. H. You, and Y.-J. Kim, *Low Frequency Acoustic Energy Harvesting Using PZT Piezoelectric Plates in a Straight Tube Resonator*, Smart Mater. Struct. **22**, 055013 (2013).
[307] B. Li, A. J. Laviage, J. H. You, and Y.-J. Kim, *Harvesting Low-Frequency Acoustic Energy Using Quarter-Wavelength Straight-Tube Acoustic Resonator*, Appl. Acoust. **74**, 1271 (2013).
[308] G. Ma, M. Yang, S. Xiao, Z. Yang, and P. Sheng, *Acoustic Metasurface with Hybrid Resonances*, Nat. Mater. **13**, 9 (2014).
[309] M. Carrara, M. R. Cacan, J. Toussaint, M. J. Leamy, M. Ruzzene, and A. Erturk, *Metamaterial-Inspired Structures and Concepts for Elastoacoustic Wave Energy Harvesting*, Smart Mater. Struct. **22**, 065004 (2013).
[310] J. Li, X. Zhou, G. Huang, and G. Hu, *Acoustic Metamaterials Capable of Both Sound Insulation and Energy Harvesting*, Smart Mater. Struct. **25**, 045013 (2016).
[311] K.-J. Ma, T. Tan, F.-R. Liu, L.-C. Zhao, W.-H. Liao, and W.-M. Zhang, *Acoustic Energy Harvesting Enhanced by Locally Resonant Metamaterials*, Smart Mater. Struct. **29**, 075025 (2020).
[312] K. H. Sun, J. E. Kim, J. Kim, and K. Song, *Sound Energy Harvesting Using a Doubly Coiled-up Acoustic Metamaterial Cavity*, Smart Mater. Struct. **26**, 075011 (2017).
[313] M. Yuan, Z. Cao, J. Luo, and Z. Pang, *Helix Structure for Low Frequency Acoustic Energy Harvesting*, Rev. Sci. Instrum. **89**, 055002 (2018).
[314] G.-S. Liu, Y.-Y. Peng, M.-H. Liu, X.-Y. Zou, and J.-C. Cheng, *Broadband Acoustic Energy Harvesting Metasurface with Coupled Helmholtz Resonators*, Appl. Phys. Lett. **113**, 153503 (2018).
[315] S.-H. Jo, H. Yoon, Y. C. Shin, M. Kim, and B. D. Youn, *Elastic Wave Localization and Harvesting Using Double Defect Modes of a Phononic Crystal*, J. Appl. Phys. **127**, 164901 (2020).
[316] M. Javadi, A. Heidari, and S. Darbari, *Realization of Enhanced Sound-Driven CNT-Based Triboelectric Nanogenerator, Utilizing Sonic Array Configuration*, Curr. Appl. Phys. **18**, 361 (2018).
[317] L. Fan, Y. He, X. Chen, and X. Zhao, *Acoustic Energy Harvesting Based on the Topological Interface Mode of 1D Phononic Crystal Tube*, Appl. Phys. Express **13**, 017004 (2019).
[318] Z. Wen, Y. Jin, P. Gao, X. Zhuang, T. Rabczuk, and B. Djafari-Rouhani, *Topological Cavities in Phononic Plates for Robust Energy Harvesting*, Mech. Syst. Signal Process. **162**, 108047 (2022).
[319] R. Lucklum and J. Li, *Phononic Crystals for Liquid Sensor Applications*, Meas. Sci. Technol. **20**, 124014 (2009).
[320] R. Lucklum, M. Ke, and M. Zubtsov, *Two-Dimensional Phononic Crystal Sensor Based on a Cavity Mode*, Sens. Actuators B Chem. **171–172**, 271 (2012).
[321] A. Oseev, M. Zubtsov, and R. Lucklum, *Gasoline Properties Determination with Phononic Crystal Cavity Sensor*, Sens. Actuators B Chem. **189**, 208 (2013).
[322] A. Salman, O. A. Kaya, and A. Cicek, *Determination of Concentration of Ethanol in Water by a Linear Waveguide in a 2-Dimensional Phononic Crystal Slab*, Sens. Actuators Phys. **208**, 50 (2014).
[323] A. Salman, O. A. Kaya, A. Cicek, and B. Ulug, *Low-Concentration Liquid Sensing by an Acoustic Mach–Zehnder Interferometer in a Two-Dimensional Phononic Crystal*, J. Phys. Appl. Phys. **48**, 255301 (2015).